\documentclass[aps,prd,twocolumn,nofootinbib,showpacs]{revtex4}

\usepackage{rotating} 
\usepackage{graphicx}
\usepackage{epsfig} 
\usepackage{amsmath} 
\usepackage{graphicx}
\usepackage{bm}

\newcommand{\zjets }{$Z$+jets}
\newcommand{\wjets }{$W$+jets}
\newcommand{\gammajets }{$\gamma$+jets}
\newcommand{\ttbarjets }{$t \bar{t}$+jets}
\newcommand{\pt }{$|\vec{p}_T|$}
\newcommand{\et }{$E_T$}
\newcommand{\ptvector}{$\vec{p}_T$}

\newcommand{\met}{$E_T\hspace{-3.9mm}/\hspace{1.5mm}$}

\newcommand{\metvector}{$\vec{E}_T\hspace{-4.0mm}/\hspace{1.6mm}$}

\newcommand{\ljetsmet}{$l$+jets+\met}
\newcommand{\vjets}{$V$+jets}

\begin{document}

\title{ Modeling missing transverse energy in $V$+jets at CERN LHC }

\author{Victor Pavlunin}
\affiliation{Department of Physics, University of California, 
	     			Santa Barbara, CA, USA, 93106-9530 }

\date{June 26, 2009}

\begin{abstract}

     I discuss a method to model the instrumental response of the  
     CMS and ATLAS detectors at high  missing transverse energy to 
     dominant standard model $V$+jets backgrounds, where $V$ is a $Z$, 
     $\gamma$ or $W$, using multi-jet QCD events. 
     The method is developed for new physics searches in early data  
     at the Large Hadron Collider~(LHC) with minimal recourse to simulation.

\end{abstract}



\maketitle

\vspace{10mm}

\section{Introduction}

The LHC enters a new energy regime to explore the origin of the 
electroweak symmetry breaking, search for and study physics beyond 
the standard model~(SM). 
At its design center-of-mass energy, new physics production 
cross sections may be significant so that
a data sample of modest integrated luminosity, 100 pb$^{-1}$ or less, 
may contain a large number of new particles.
The challenge is to distinguish events 
with new particles from those, many orders of magnitude more copious, 
attributed to the SM with limited understanding of the SM production 
rates and detector performance in early LHC data.

Missing  transverse energy, \met, has discriminating 
power to reveal new particles interacting weakly with ordinary matter 
produced via high energy parton collisions in laboratory conditions at the LHC. 
These weakly interacting particles may comprise the dark matter 
of our Universe~\cite{dark-matter}. 
They are expected in new physics models, 
such as R-parity conserving super-symmetry~\cite{susy} 
and many others~\cite{other-models}. 
Missing transverse energy allows to perform a broad 
search sensitive to the presence of such particles in collision data and  
is an observable that may lead to an early discovery at the LHC~\cite{extra-d}. 
At the same time, missing transverse energy is one of the most 
difficult observables to measure precisely and simulate 
accurately~\cite{met-effects}
because it is measured by multiple detector sub-systems and 
subject to mis-measurements and backgrounds in any of them.

In this paper, I discuss a new method~\cite{previous-methods} to predict 
backgrounds at high \met~ for new physics searches in signatures consistent with 
SM $V$+jets~ and \ttbarjets~\cite{ttbar}, where $V$ is a $Z$, $\gamma$ or $W$. 
I assume that new particles are heavy and decay to SM particles 
emitting multiple jets so that high sensitivity is expected 
at high \met~ and a large number of jets. Since the main sources of artificial 
\met~ come from the system of jets, the detector and non-collision effects,
I model the instrumental response to the system of jets in $V$+jets and 
other effects at high \met~ {\it in-situ} using multi-jet QCD events. 
This method complements and extends the work of Ref.~\cite{forward-modeling}, 
where events with high rapidity objects are used to model 
SM $V$+jets and multi-jet backgrounds in new physics searches 
without heavy reliance on \met. The emphasis of this work, 
as that of Ref.~\cite{forward-modeling}, 
is on robustness against imperfections of background 
modeling required for new physics searches in early LHC data.

\section{Overview}

\label{method}

Monte Carlo~(MC) simulation capable of modeling the detector response 
to SM processes is a great asset in new physics searches. 
However, there are two challenges in searches of early LHC data 
based on MC simulation.
First, the SM $V+$jets production rates are difficult to predict 
from first principles. 
MC techniques are unreliable in predicting backgrounds with a large 
number of jets and need to be tuned with high $\sqrt{s}$ data.
Theory calculations at sufficiently high order 
in many cases do not exist~\cite{theory_predictions}. 
The structure functions have significant uncertainties 
in the small $x$ range accessible at the LHC~\cite{pdf_uncert}. 
Second, significant uncertainties in the calibration 
of the experimental apparatus are expected 
in early data taking. 
Missing transverse energy is an observable that is particularly
difficult to measure precisely and simulate accurately, 
since large jet energy fluctuations, detector artifacts, 
collision related and non-collision effects can produce 
non-Gaussian high \met~ tails. 
These artificial \met~ tails may resemble a signature of 
a new weakly interacting particle.

To introduce the method, let us consider an event with a $Z$ boson 
reconstructed in the di-muon channel and four jets. The four-momentum 
of the $Z$ is well-measured so that the system of the four jets and 
other effects unrelated to the di-muon system are the main source of 
\met~ in this event. 
In order to develop a search in \met~ based on MC simulation, 
one would need to identify, understand and simulate the detector 
response to each of these effects. 
Instead, I model these effects in-situ 
using multi-jet QCD events as follows. 
A sample of QCD events with four jets that have approximately the 
same configuration as the four jet system of the \zjets~ event is selected.
A \met~ prediction, or a template, for this \zjets~ event is obtained 
using the \met~ distribution measured in the selected QCD sample  and
normalized to unity. This procedure is repeated for other \zjets~  
events with four jets. The \met~ templates are summed to 
obtain a SM \met~ prediction for all \zjets~ events with 
four jets.

The photon momentum in \gammajets~ is also well-measured so that the same 
procedure applies to obtain a SM \met~ prediction in the \gammajets~ sample. 
In \wjets~ and \ttbarjets~ with one of the two top quarks decaying 
semileptonically, the \ljetsmet~ signature, there is genuine \met~ from 
the undetected neutrino produced in $W$ decays. 
To avoid reliance on MC and theory, 
I model the neutrino \pt~ spectra using the charged lepton \pt~ spectra. 
If the $W$ bosons are not polarized in the transverse plane, the two spectra should be the same. 
Event selection and polarized $W$ bosons produced in top quark decays 
lead to differences in the charged lepton and neutrino spectra.
However, these differences are small and can be accounted for by corrections. 
A prediction for artificial \met~ in \wjets~ and \ttbarjets~ is derived in the 
same manner as for \zjets~ and combined with a modeled neutrino \pt~ spectrum
to obtain a SM \met~ prediction in the \ljetsmet~ final state.

In each channel, the search is made in \met~ distributions of
events with the same number of jets, or events with at least 
a certain number jets. Since higher sensitivity to new physics is expected 
in events with a large number of jets, the focus is to model 
the high \met~ region in \vjets~ events with 3 or more jets. 
Events with 2~jets are valuable as a validation and calibration 
sample. This method is developed for searches in early LHC data.
It will work best if the LHC start-up is quick, new particles are strongly 
produced and not very heavy, $e.g.,$ such as squarks and gluinos in the 
low mass mSUGRA CMS and ATLAS benchmarks~\cite{previous-methods}. 
With this in mind, a prediction of SM backgrounds in high \met~ tails 
to about 20\% may be sufficient to reveal new physics. 
For this reason, an accuracy benchmark for this method is to predict SM 
backgrounds in $V$+jets at high \met~ and a large number of jets~(3 jets 
or more) to about 20\% or better in a data-driven manner.

\section{Experimental Aspects}

\label{experiment}

The CMS and ATLAS experiments use multi-purpose detectors at the European 
Organization for Nuclear Research~(CERN). 
Detailed descriptions of the detectors can be found in Ref.~\cite{detectors}. 
The detectors are capable of reconstructing electrons and muons with high 
efficiencies and low fake rates for lepton \pt$\;> 20$~GeV in the $|\eta| < 2.5$ range~\cite{eta}. 
(In this paper the symbol~$l$ is used to denote 
an $e$ or $\mu$, but not $\tau$. Charge-conjugate modes are implied.) 
In both detectors, photons and jets can be reconstructed 
reliably within $|\eta| < 2.5$ and $|\eta| < 3.0$, respectively. 

To study the method, mock data samples are generated for 
the following SM processes: \zjets~(5.0~fb$^{-1}$, up to 5 partons, 
$Z \rightarrow l^+ l^-$), \wjets~(1.0~fb$^{-1}$, up to 5 partons, $W \rightarrow l \nu_l$), 
\ttbarjets~(1.0~fb$^{-1}$, up to 4 partons, $t\bar{t} \rightarrow l \nu_l bbjj$), 
\gammajets~(400.0~pb$^{-1}$, up to 5 partons) and QCD jets 
(1.0~pb$^{-1}$, up to 5 partons)~\cite{sqrts-at-LHC}.
(The same samples are used in Ref.~\cite{forward-modeling}.) 
The integrated luminosity listed in parentheses is used everywhere 
in tests in this paper but section~\ref{prediction-with-signal}. 
These samples are generated with ALPGEN~\cite{alpgen} at the parton level. 
PYTHIA~\cite{pythia} is used for parton showering, hadronization, simulation 
of the underlying event and jet reconstruction.
To model features of a new physics signal in search distributions, 
mock signal data samples for Minimal Supergravity (mSUGRA) 
benchmark points LM1 and LM4~\cite{susy,msugra,cms-msugra} 
are generated with PYTHIA.

Electrons and muons are required to have \pt~ of at least 20~GeV 
in the $|\eta| < 2.5$ range. Photons are reconstructed above 
the \pt~ threshold of 30~GeV in the $|\eta| < 2.5$ range. 
Jets are reconstructed using the PYCELL algorithm for 
$R=0.5$~\cite{pythia} and 
required to be within $|\eta| < 3.0$.
A low jet \pt~ threshold of 20~GeV is used in the \met~ measurements 
in order to collect the energy deposited in the calorimeters to 
a fuller extent. Higher jet \pt~ thresholds, 50 GeV or more, 
are used to measure other observables in a robust manner 
as indicated below. 
I require that the leading jet~ and \metvector~ 
be not aligned in the transverse plane within 0.15~radians:
$0.15 < | \Delta\phi^{\rm lead\; jet-{\it E}_T\hspace{-3.5mm}/\hspace{1.5mm}} | < (\pi - 0.15)$.
(The jet with the highest \pt~ in an event is the leading jet 
of this event. Any other jet in this event is a non-leading jet.)
It is assumed that the triggering and event reconstruction 
efficiency in each channel is 50\%.

\begin{figure}[h!]
\begin{center}
\epsfig{file=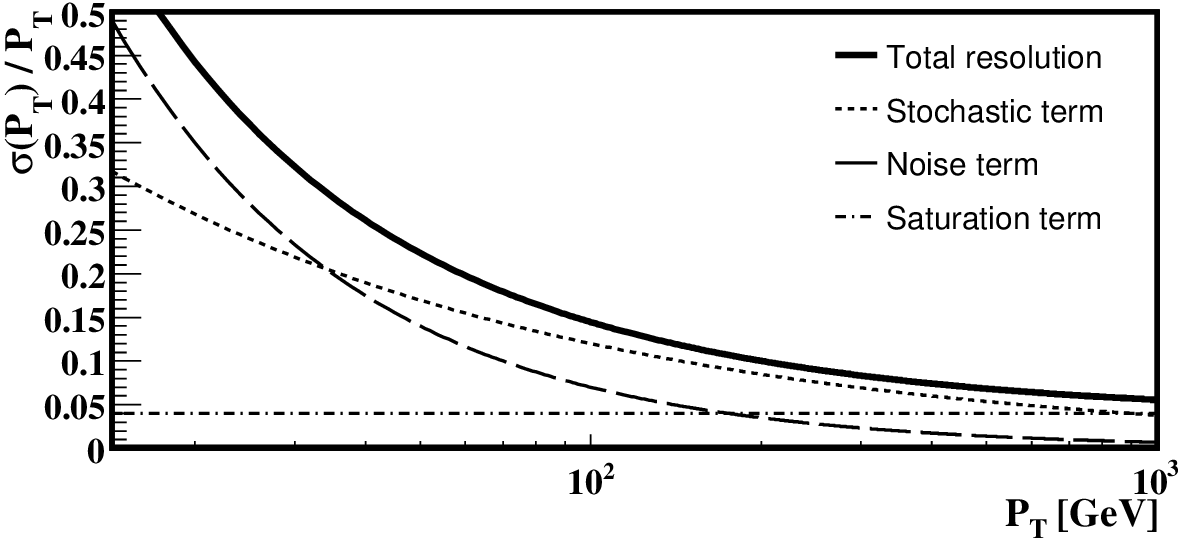,width=3.0in}\\
\epsfig{file=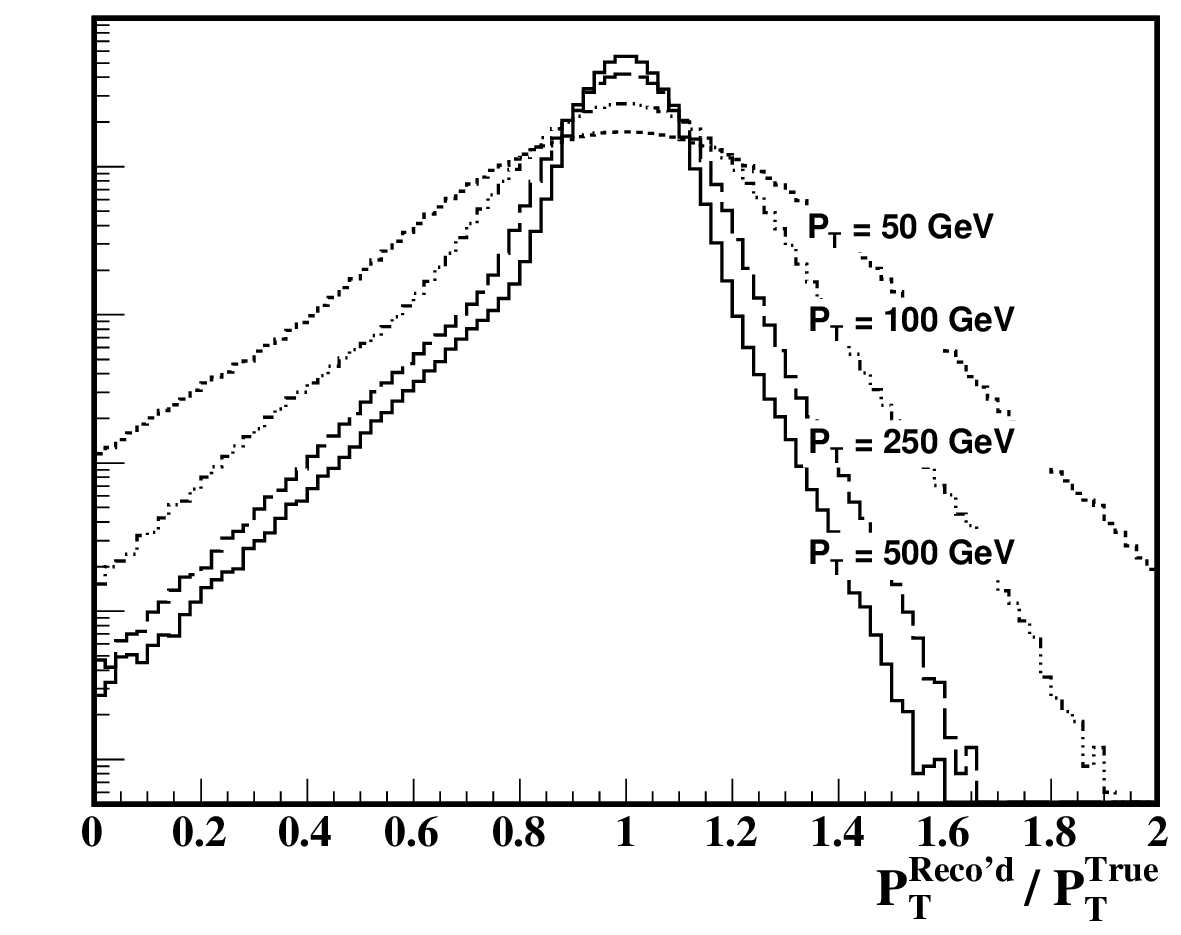,width=3.0in}
\end{center}
\caption{ 
Top: the Gaussian contribution to the jet energy 
resolution as a function of true jet \pt~ is shown 
in the solid line. 
The noise, stochastic and saturation contributions 
to the jet resolution function are shown separately. 
Bottom: the jet energy smearing functions for 500, 250, 
100 and 50~GeV \pt~ jets are shown in the solid, dashed, 
dot-dashed and dotted lines. }
\label{figure-1}  
\end{figure}

The vector of missing transverse energy, \metvector, is calculated as 
the vector opposite to the sum of $\vec{p}_T$ measurements of charged 
leptons, photons, and jets in each event.   
Since electron, muon and photon momenta are measured with high precision 
compared to jets, their contribution to artificial \met~ in events with 
a large number of jets is negligible.
To emulate detector resolution effects for jets, 
jet energies measured by PYTHIA are smeared.  
The jet smearing function has three components: 
a) a Gaussian with 
\[
\sigma{(|\vec{p}_T|)}/|\vec{p}_T| = \sqrt{(7.0/|\vec{p}_T|)^2+(1.2/\sqrt{|\vec{p}_T|})^2+(0.04)^2},
\]
where $|\vec{p}_T|$ is measured in GeV, 
b) an exponential low-side tail 
that stretches from 
$[1.0 - 2 {\sigma(|\vec{p}_T|)}/{|\vec{p}_T|}]$ to 0.0  
added to the Gaussian component 3\% of the time, 
and c) similarly, an exponential high-side tail  
from $[1.0 + 2 {\sigma(|\vec{p}_T|)}/{|\vec{p}_T|}]$ to infinity 
added to the Gaussian 1\% of the time. 
Figure~\ref{figure-1} shows the \pt~ dependence of the Gaussian 
smearing and the full smearing function with the non-Gaussian tails  
for a few fixed jet \pt~ values. 
This jet smearing function is constructed based on 
studies of the CMS and ATLAS detectors~\cite{previous-methods,detectors} 
to represent the expected jet response characteristic of the two detectors. 
Since the jet system tends to be the dominant source of 
artificial \met, all three components of the jet smearing function 
are varied in the studies of robustness and limitations of 
the method discussed below.

The selection criteria used in the paper are not optimized to any new 
physics model. Instead, they are chosen to ensure robust detector 
performance and maintain sensitivity to a wide range of new physics 
models at high \met~ and a large number of jets.  
The selection criteria can be modified without producing
a significant effect on  the method's performance. 
The mSUGRA benchmarks listed above are 
used only to illustrate generic features of a new physics contribution. 
The goal of this paper is to demonstrate the scope of 
the method and its performance rather than to attain high 
sensitivity to a specific model or to quantify it.

\section{ Algorithm }

\label{algorithm}

I will describe in detail the algorithm in this section and 
present results of its tests in the next section.
Let us assume that the $V$~momentum in a $V$+jets event is known. 
The resolution and other effects producing artificial missing transverse 
energy for this event are modeled using multi-jet QCD events 
with a kinematic configuration of jets similar to that in 
the $V$+jets event.

A prominent difference between the jet systems in $V$+jets and 
QCD is that in $V$+jets events the jet system recoils against 
the $V$ while  in QCD events it is at rest in the transverse plane.
A key is to select 
multi-jet QCD events to predict \met~ in a manner that captures 
effects generating artificial \met~ in-situ but allows for the difference stemming from 
the boost of the jet system in the transverse plane in $V$+jets.

It is not necessary to model accurately every degree of freedom in 
the jet system of $V+$jets on an event-by-event basis 
by QCD for two reasons.
First, each $V+$jets event is modeled by a large sample of QCD 
events so that mis-modeled correlations are averaged out over
this sample of QCD events. 
Second, \met~ is measured for the entire $V$+jets sample so 
that mis-modeled correlations are averaged out over $V$+jets events 
as well. These averaging effects allow to develop a simple algorithm.

Multi-jet QCD events are selected using two variables: 
(1)~$N_J$, number of jets above a high \pt~ threshold 
(50~GeV or higher), and 
(2)~$J_T \equiv \sum |\vec{p}_{T}^{\rm \;jet}|$ for
jets above a low 20~GeV \pt~ threshold (the same jet threshold 
is used for \met~ measurements)~\cite{elec-jet-ambiguity}.   
A QCD sample is selected for each pair ($N_J$, $J_T$~bin),
the width of $J_T$~bins is 10~GeV~(100~GeV) below~(above) 1~TeV.
A \met~ template is obtained for each of these samples 
as a \met~  distribution in that sample normalized to unity. 
For each $V$+jets event, $N_J$ and $J_{T}$ are measured and 
used to select the \met~ template with the same $N_J$ in the 
corresponding $J_T$ bin, 
which represents an artificial \met~ prediction for this $V$+jets event. 
These templates can be summed over, for example, all $V$+jets 
events to obtain a \met~ prediction for the entire $V$+jets sample.

Two sets of jet \pt~ thresholds are used to measure $N_J$. In the first set, 
the jet \pt~ threshold for $N_J$ is 50~GeV. QCD events for this jet 
threshold can be collected using prescaled low \pt~ jet triggers. 
In the second set, the jet thresholds are equal to (or higher than) 
the jet \pt~ thresholds that can be used in unprescaled multi-jet triggers. 
For the second set, 
I use 140~GeV for $N_J = 2$,  80~GeV for $N_J = 3$ and 60~GeV for $N_J \ge 4$.
These jet \pt~ thresholds can be changed depending 
on the trigger rates in data without significant effect.

\begin{figure*}[pt!]
\begin{center}
\begin{minipage}{7.1in}
\begin{center}
\epsfig{file=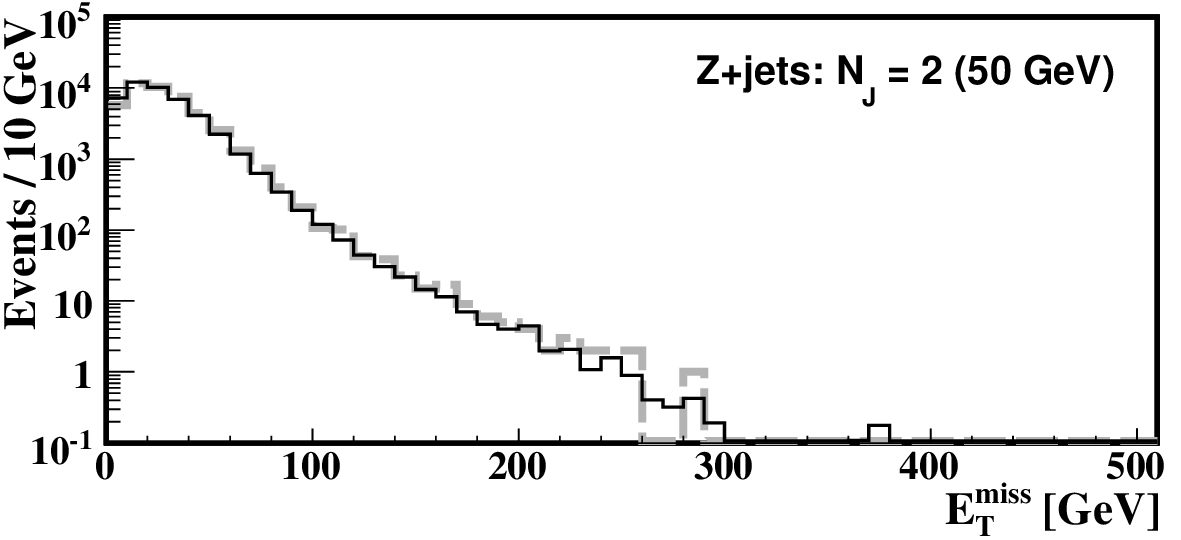,width=2.8in}
\epsfig{file=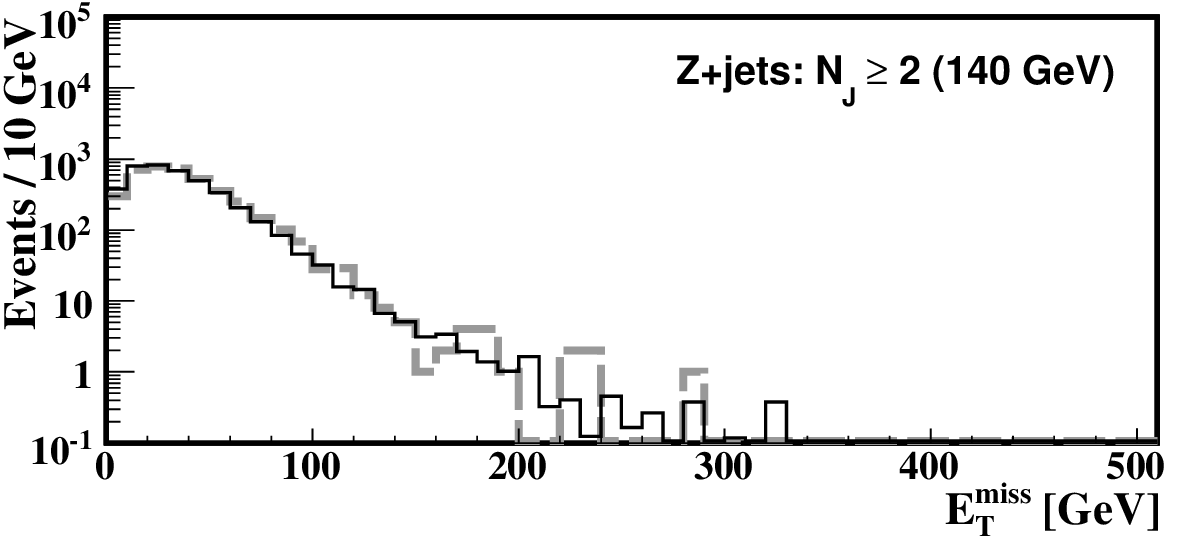,width=2.8in}
\end{center}
\end{minipage}
\begin{minipage}{7.1in}
\begin{center}
\epsfig{file=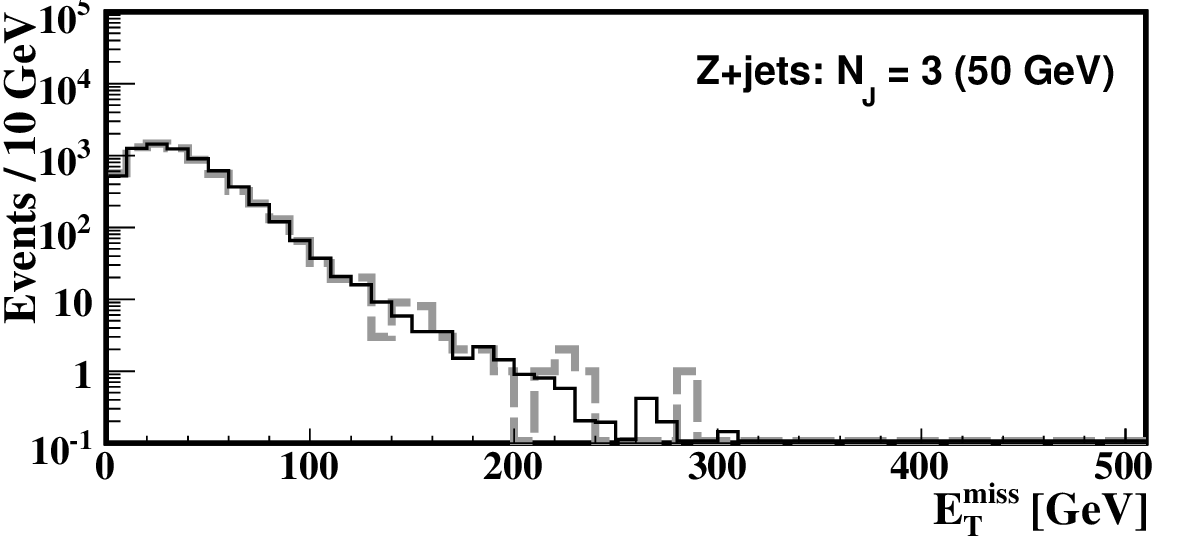,width=2.8in}
\epsfig{file=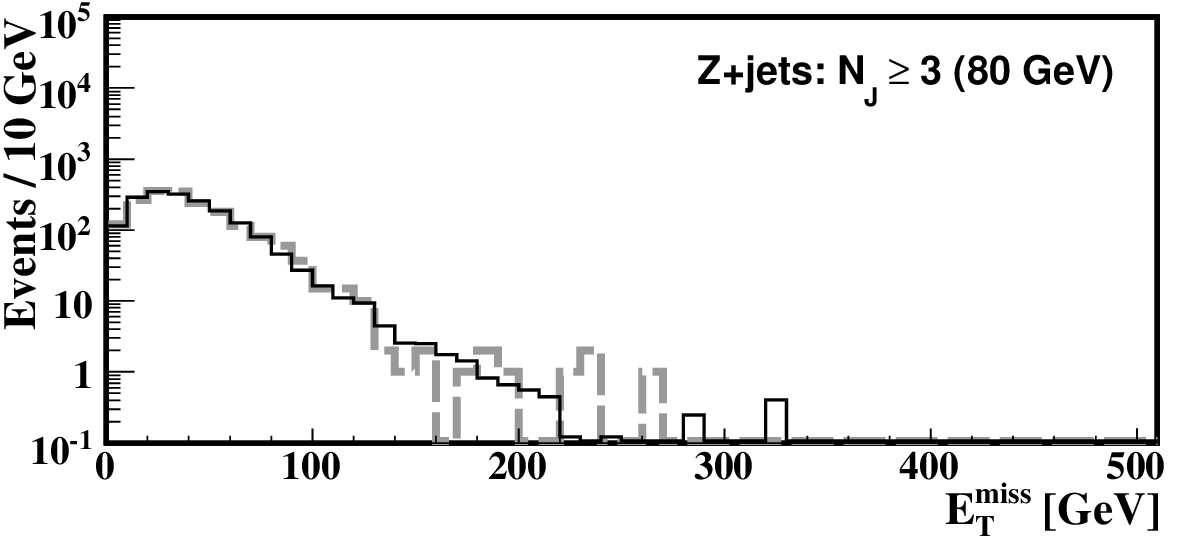,width=2.8in}
\end{center}
\end{minipage}
\begin{minipage}{7.1in}
\begin{center}
\epsfig{file=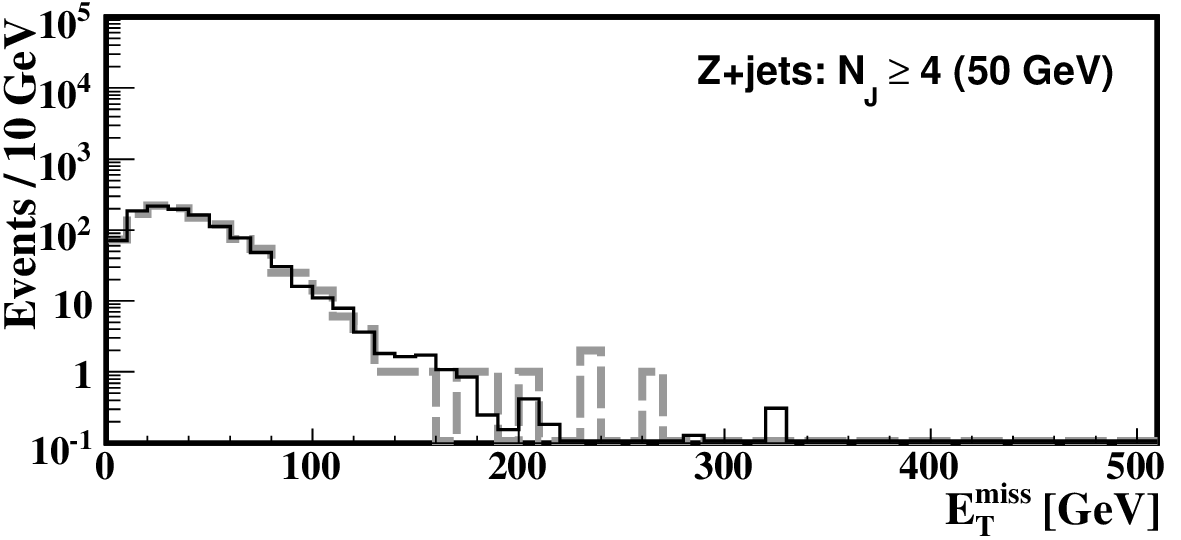,width=2.8in}
\epsfig{file=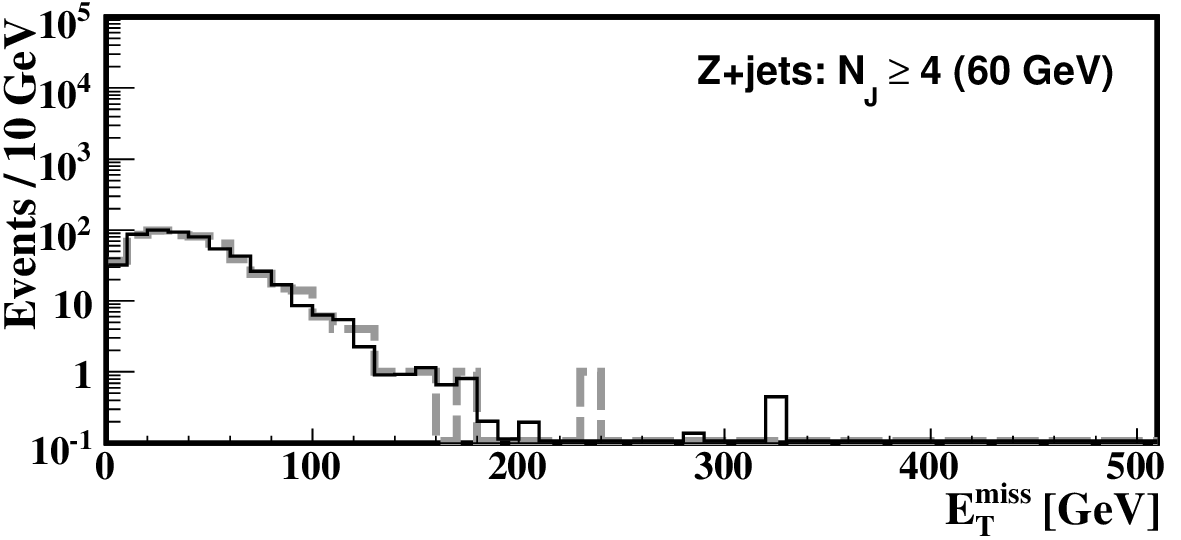,width=2.8in}
\end{center}
\end{minipage}
\end{center}
\caption{ Algorithm performance in \zjets~ for $N_J$ of, or at least of, 2, 3 and 4 in the first, middle 
	  and third rows, respectively. 
          The first~(second) column is for the the 50~GeV~(high) jet \pt~ thresholds for $N_J$.
          The observed \met~ distributions are shown in the dashed lines, 
          their predictions obtained using multi-jet QCD events are the solid lines. }
\label{figure-2}
\begin{center}
\begin{minipage}{7.1in}
\begin{center}
\epsfig{file=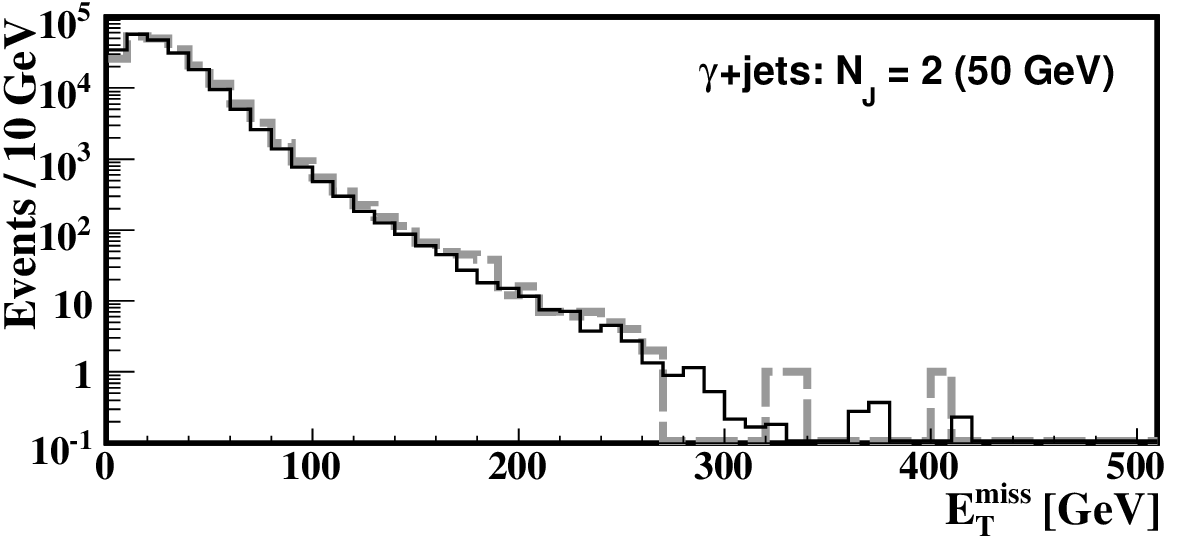,width=2.8in}
\epsfig{file=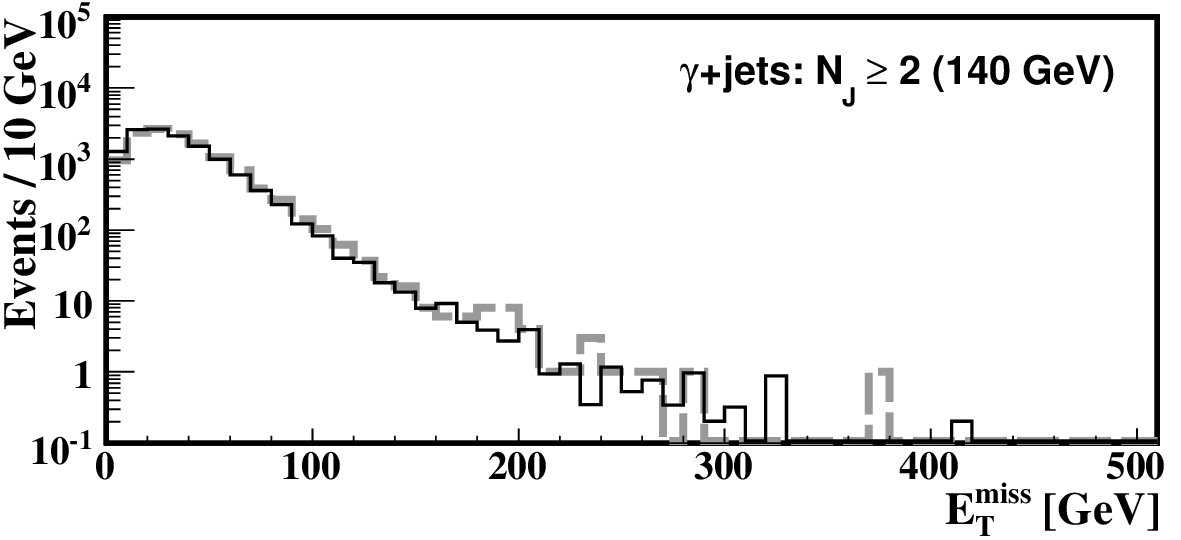,width=2.8in}
\end{center}
\end{minipage}
\begin{minipage}{7.1in}
\begin{center}
\epsfig{file=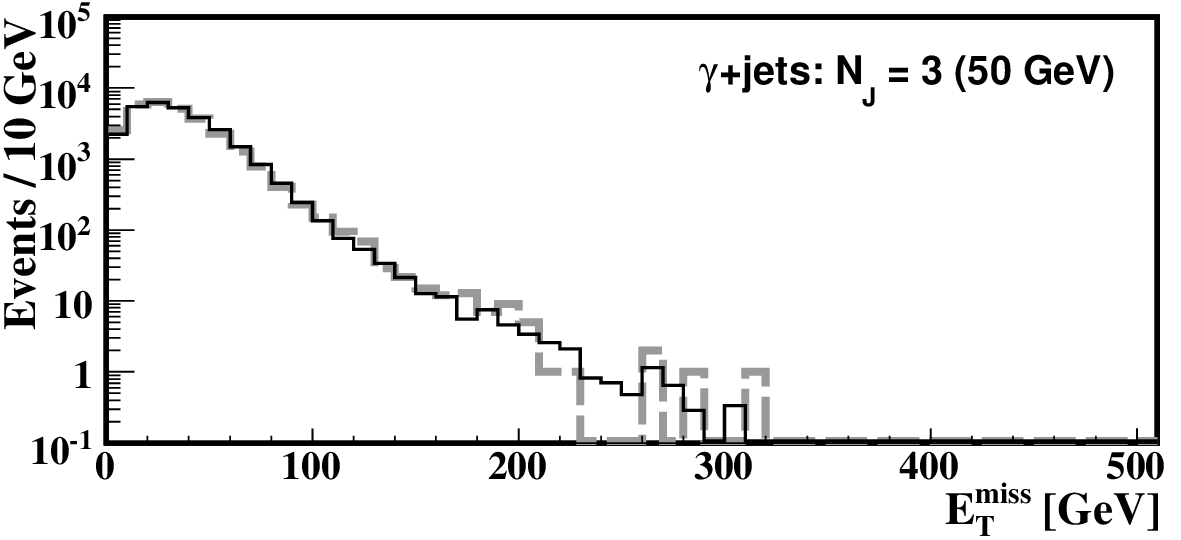,width=2.8in}
\epsfig{file=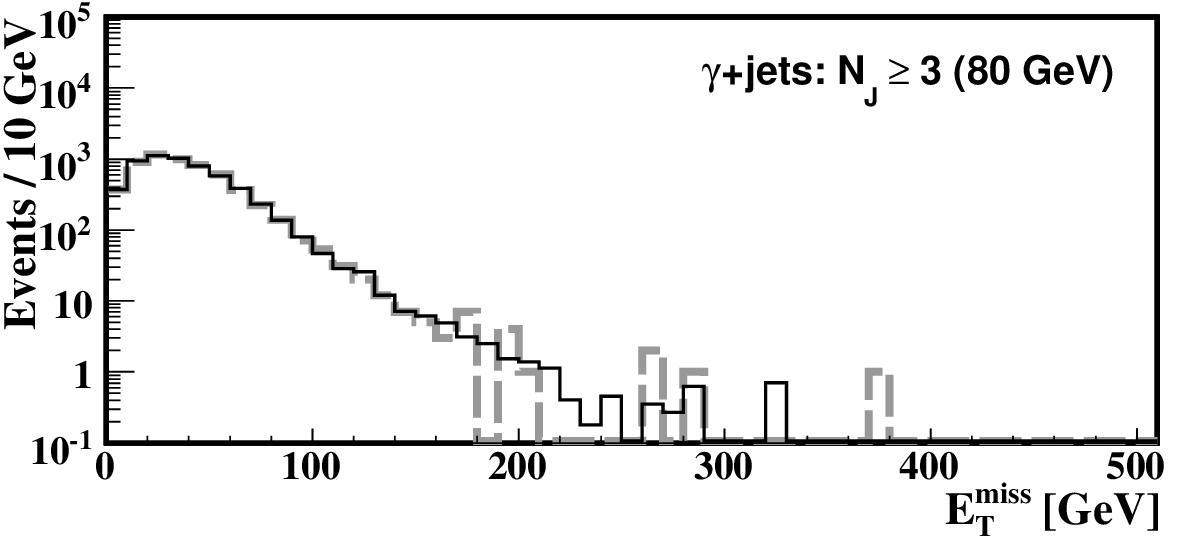,width=2.8in}
\end{center}
\end{minipage}
\begin{minipage}{7.1in}
\begin{center}
\epsfig{file=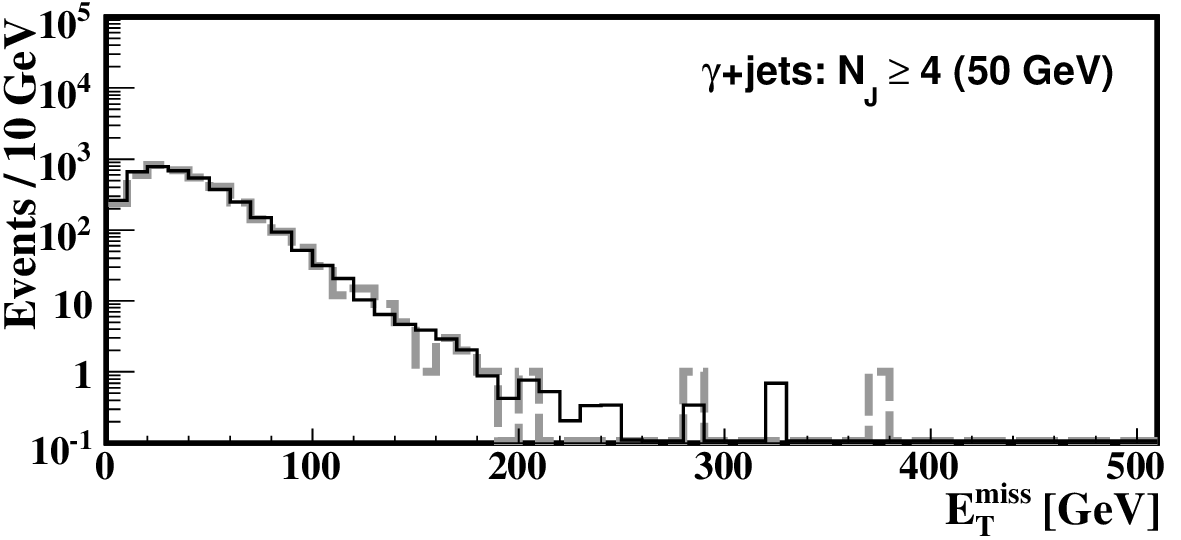,width=2.8in}
\epsfig{file=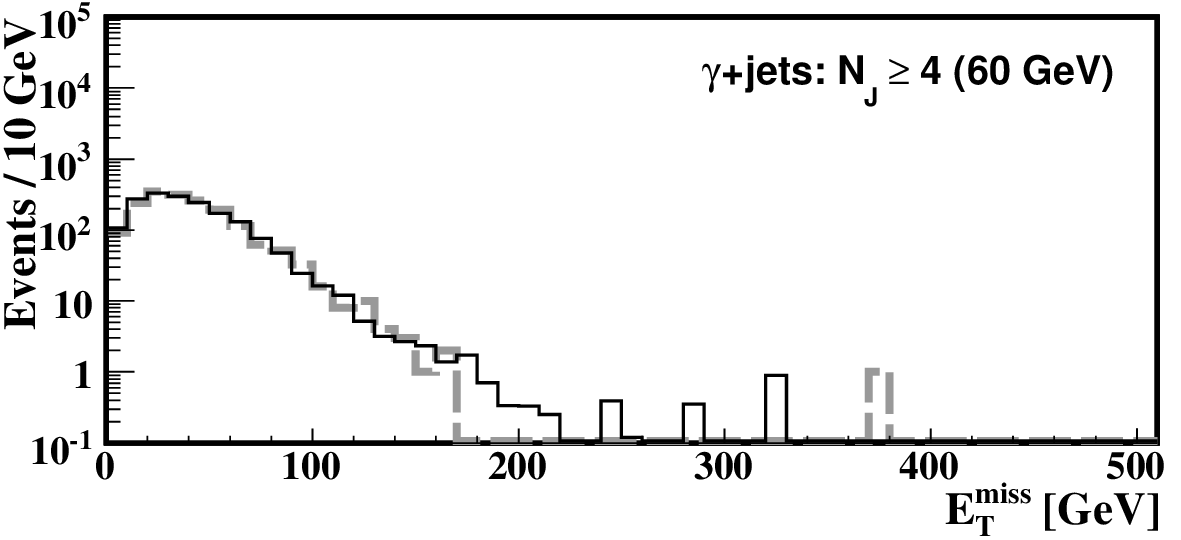,width=2.8in}
\end{center}
\end{minipage}
\end{center}
\caption{ 
          Algorithm performance in \gammajets~ for  $N_J$ of, or at least of, 2, 3 and 4 in the first, middle 
	  and third rows. 
          The first~(second) column shows results for the 50~GeV~(high) jet \pt~ threshold(s) for $N_J$.
          The observed \met~ distributions are shown in the dashed lines, 
          their predictions are the solid lines.
	}
\label{figure-3}
\end{figure*}

The $V$ momentum in \zjets~ and \gammajets~ is well-measured 
so that the application of the algorithm is straightforward in these channels. 
A comparison of predicted and observed yields is shown in 
Figures~\ref{figure-2} and~\ref{figure-3} for \zjets~ and \gammajets,
respectively, where the top~(bottom) row shows results for the 50~GeV~(high) 
jet \pt~ thresholds for $N_J$. It is seen that the prediction~(solid line) 
is very good for $N_J = 3$~(50~GeV threshold), $N_J \ge 3$~(high threshold)  
and $N_J \ge 4$~(50~GeV and high thresholds). 
For events with $N_J = 2$~(50~GeV threshold) or $N_J \ge 2$~(high threshold), 
the measurement~(dashed line) is about 20\% below the prediction at \met$\; \le 20$~GeV. 
The mechanism responsible for this bias is discussed in section~\ref{limitations}. 
Since new physics is not expected to contribute at small \met, 
to remove this bias, the prediction is normalized to the measurement 
in the \met~$\in [50,\;100]$~GeV interval. 
This is done for the \met~ predictions in \zjets~ and \gammajets~ events 
with $N_J = 2$~(50~GeV threshold) or $N_J \ge 2$~(high threshold) 
everywhere in the rest of the paper.

\begin{figure*}[pt!]
\begin{center}
\begin{minipage}{7.1in}
\begin{center}
\epsfig{file=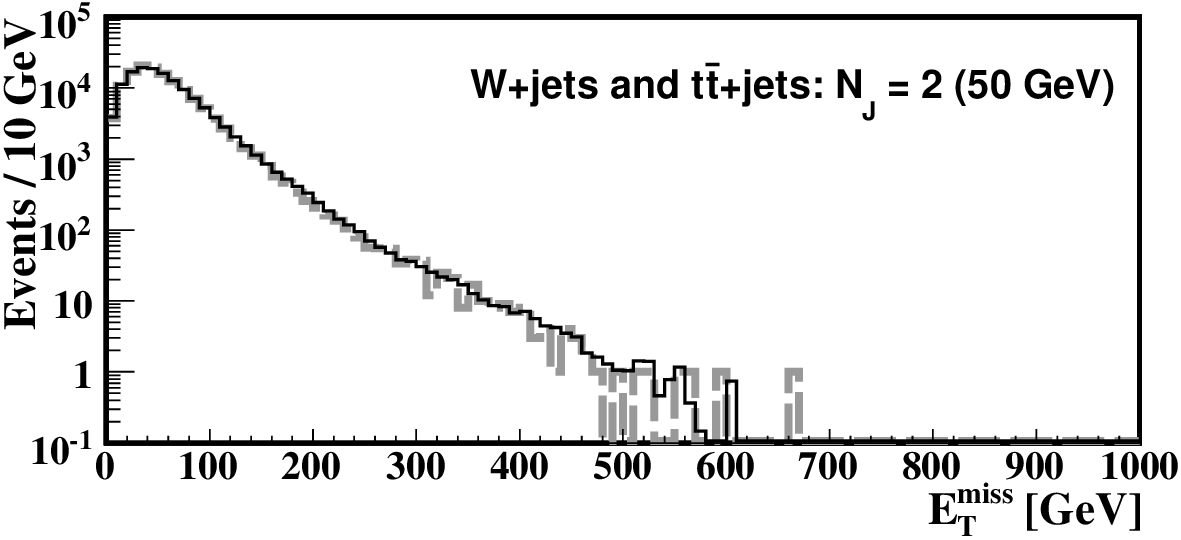,width=2.8in}
\epsfig{file=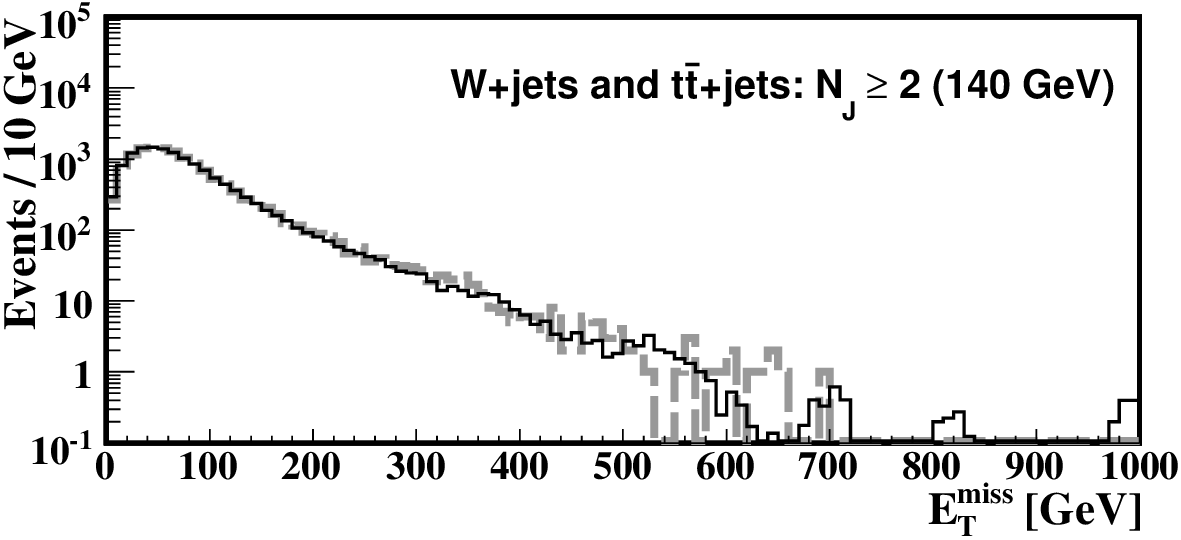,width=2.8in}
\end{center}
\end{minipage}
\begin{minipage}{7.1in}
\begin{center}
\epsfig{file=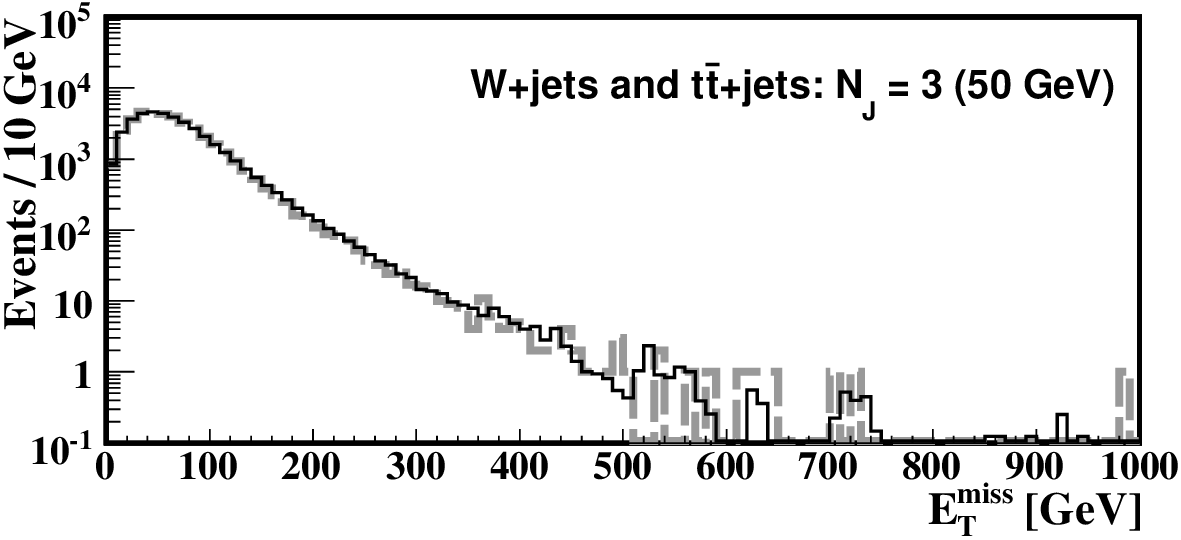,width=2.8in}
\epsfig{file=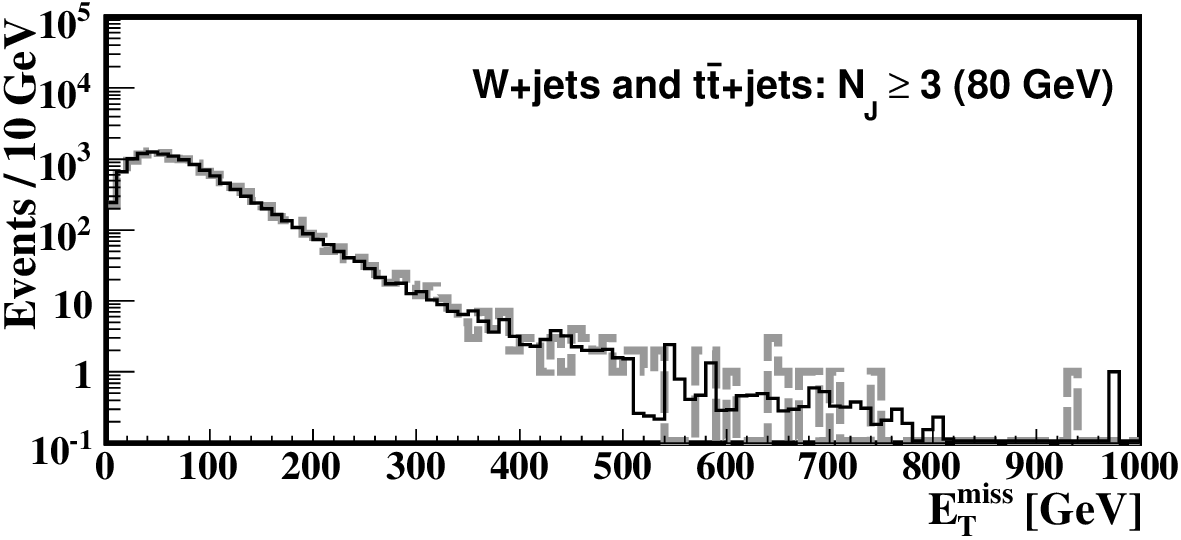,width=2.8in}
\end{center}
\end{minipage}
\begin{minipage}{7.1in}
\begin{center}
\epsfig{file=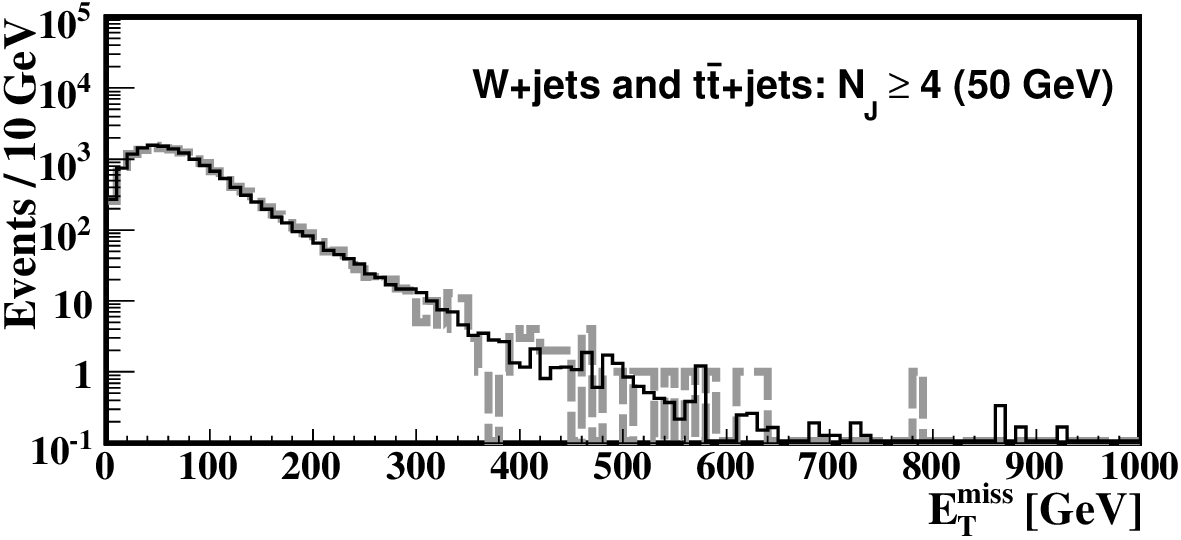,width=2.8in}
\epsfig{file=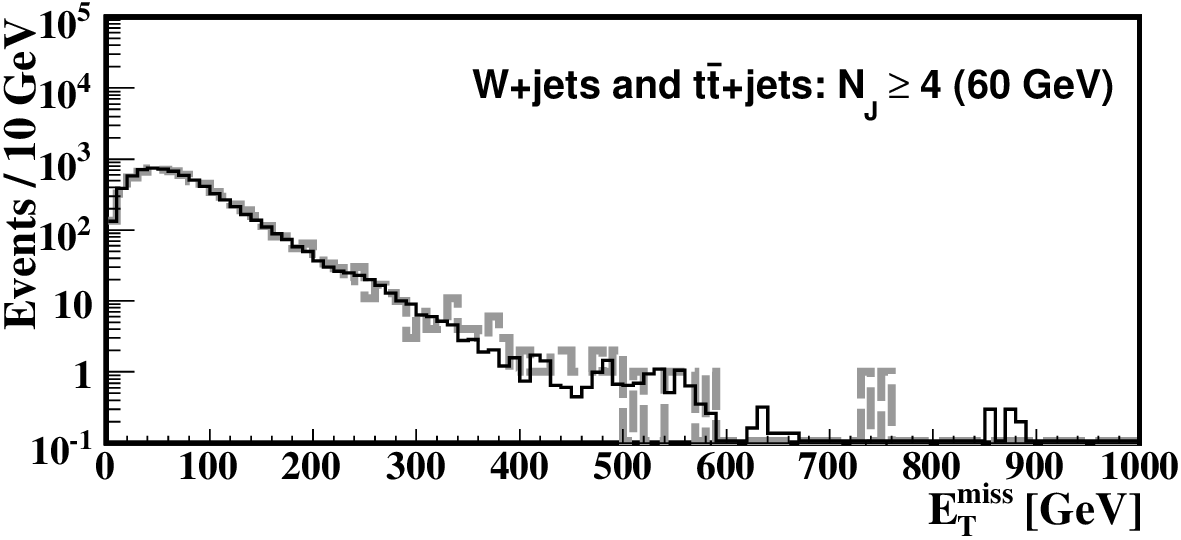,width=2.8in}
\end{center}
\end{minipage}
\end{center}
\caption{
	  Algorithm performance in  \ljetsmet~ for $N_J$ of, or at least of, 2, 3 and 4 in the first, middle 
	  and third rows. 
          The first~(second) column shows results for the 50~GeV~(high) jet \pt~ threshold(s) for $N_J$.
          The observed \met~ distributions are shown in the dashed lines, 
          their predictions are the solid lines.
	}
\label{figure-4}
\end{figure*}

In \wjets~ and \ttbarjets~ events reconstructed in the $l$+jets+\met~ channel, 
there is genuine missing transverse energy from undetected neutrinos 
produced in $W$ decays. 
Initially, to study only the effect of \met~ mis-measurements,
I consider the dominant $W \rightarrow l \nu_l$ and 
$t\bar{t} \rightarrow l \nu_l b\bar{b}jj$ contributions and 
assume that the neutrino \pt~ spectra are known until section~\ref{nu-spectrum}.
To model \met~ resolution effects, the neutrino \pt's are smeared 
with the artificial \met~ predictions obtained from multi-jet QCD events. 
This is done on an event-by-event basis assuming that the neutrino \ptvector~ and 
the artificial \metvector~ interfere at a random angle $\phi$ distributed 
uniformly from 0 to $\pi$ in the transverse plane.
Figure~\ref{figure-4} shows how well the method works  
for the 50~GeV~(first column) and high~(second column) jet \pt~ thresholds 
for $N_J$ in the $l$+jets+\met~ final state, where both \wjets~ 
and \ttbarjets~ are included according to their expected
production cross sections.

In section~\ref{nu-spectrum}, it is demonstrated that one can 
approximate the neutrino \pt~ spectra by the charged lepton \pt~ spectra. 
The contribution from $W \rightarrow \tau \nu_\tau$ in \wjets~ and \ttbarjets~ 
is also considered in section~\ref{corrections}.  With these extensions, 
the method can be used to predict the \met~ distribution in 
the $l$+jets+\met~ final state, which has high sensitivity to 
a variety of new physics models with new weakly interacting 
particles in early data.

\begin{figure}[ht!]
\begin{center}
\epsfig{file=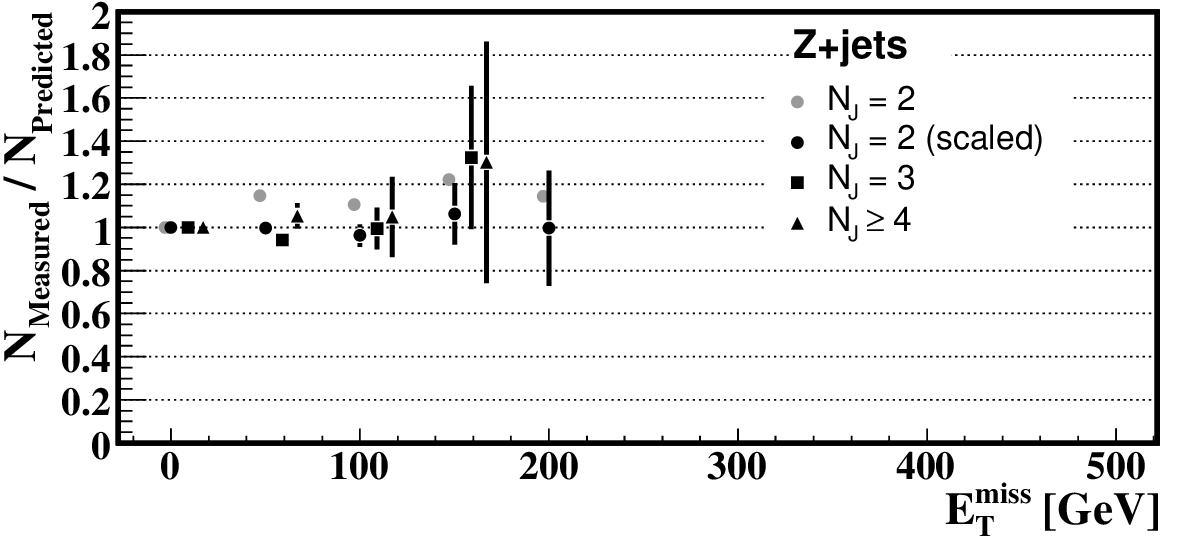,width=3.5in}\\
\epsfig{file=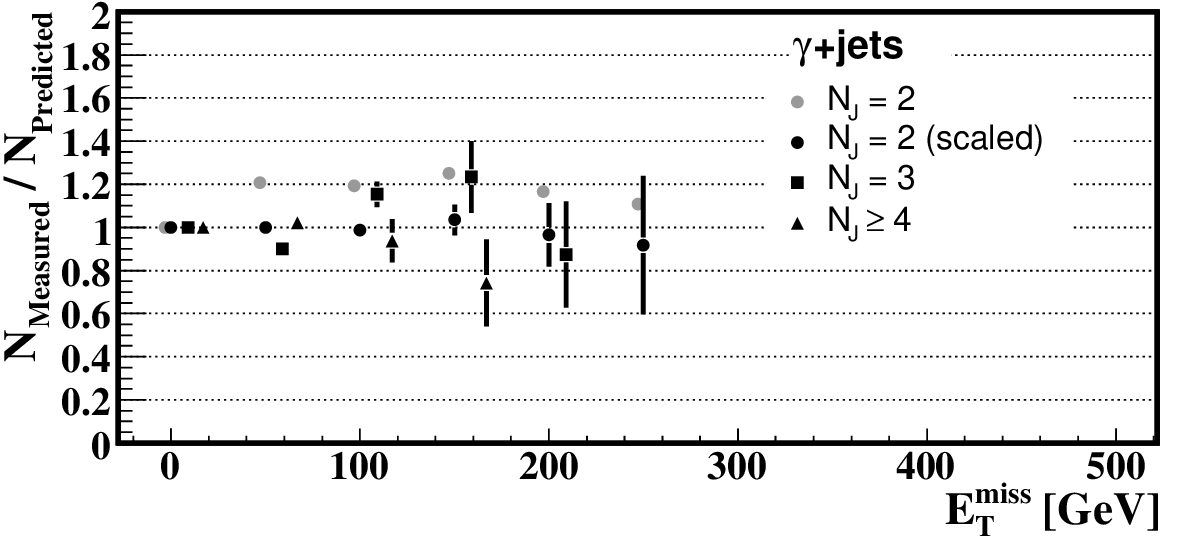,width=3.5in}\\
\epsfig{file=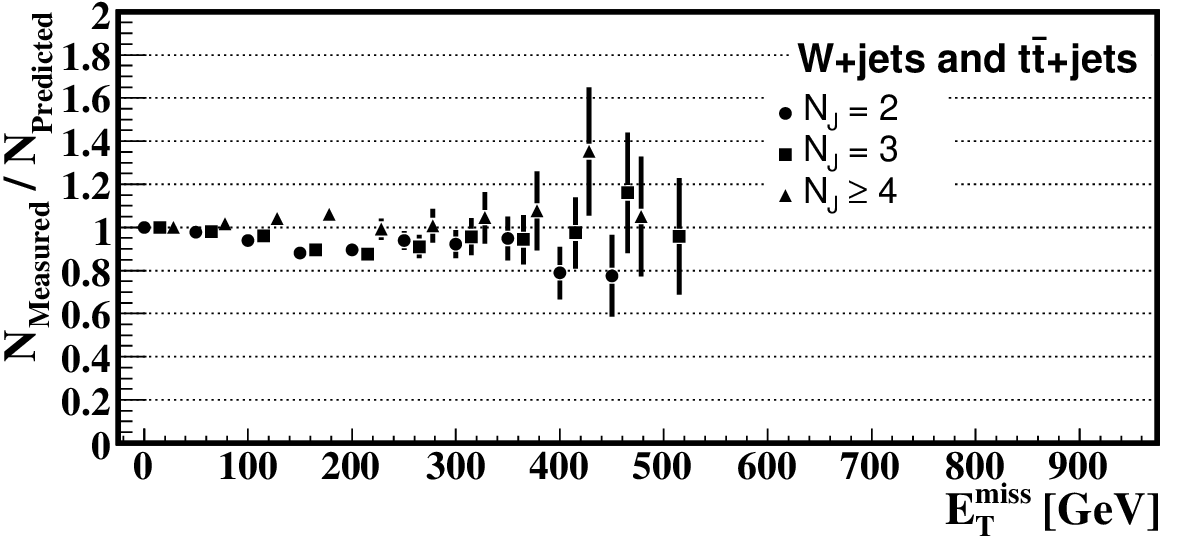,width=3.5in}
\end{center}
\caption{ Ratios of observed and estimated integrated yields
          for \zjets~(top), \gammajets~(middle),  
          $l+$jets+\met~(bottom) obtained for the 50~GeV jet \pt~ threshold
          for $N_J$. 
	  In each plot three types of markers are shown for  
          $N_J = 2$~(circles), 3~(squares) and $\ge 4$~(triangles).  
          The shaded markers for \zjets~ and \gammajets~
          show the ratios before the predictions
          are normalized at low \met~ as described 
          in the text. 
          Note, the ratios are correlated since yields are integrated upwards.}
\label{figure-5}
\end{figure}

For brevity, in the rest of the paper, I present results of 
studies for the 50~GeV jet threshold used to measure $N_J$. 
They have higher statistical precision than those for higher jet \pt~  
thresholds for $N_J$. Ratios of observed and predicted yields, 
$N_{\rm Observed}/N_{\rm Predicted}$, are shown in Figure~\ref{figure-5}, 
where the yields in each \met~ bin are integrals of the distributions shown 
in the first columns of Figures~\ref{figure-2} through~\ref{figure-4} from that 
bin's \met~ value to infinity. 
The algorithm performs at least as well when the set 
of higher jet \pt~ thresholds for $N_J$ is used.

Since the QCD production cross section is very large at the LHC, only a small QCD 
sample is needed for this method to work, $e.g.,$ 1~pb$^{-1}$ of QCD is used to 
model \met~ distributions in 5~fb$^{-1}$ of \zjets~ in this paper. 
Again, the QCD sample for templates can be collected via prescaled 
small \pt~ jet  triggers and unprescaled high \pt~ multi-jet  triggers. 
Due to the large QCD production cross section, the relative contribution 
from electroweak processes with genuine \met~ from neutrinos in this 
sample is negligible for searches in early data.

\section{Robustness}

\label{robustness}

\begin{figure}[h!]
\begin{center}
   \epsfig{file=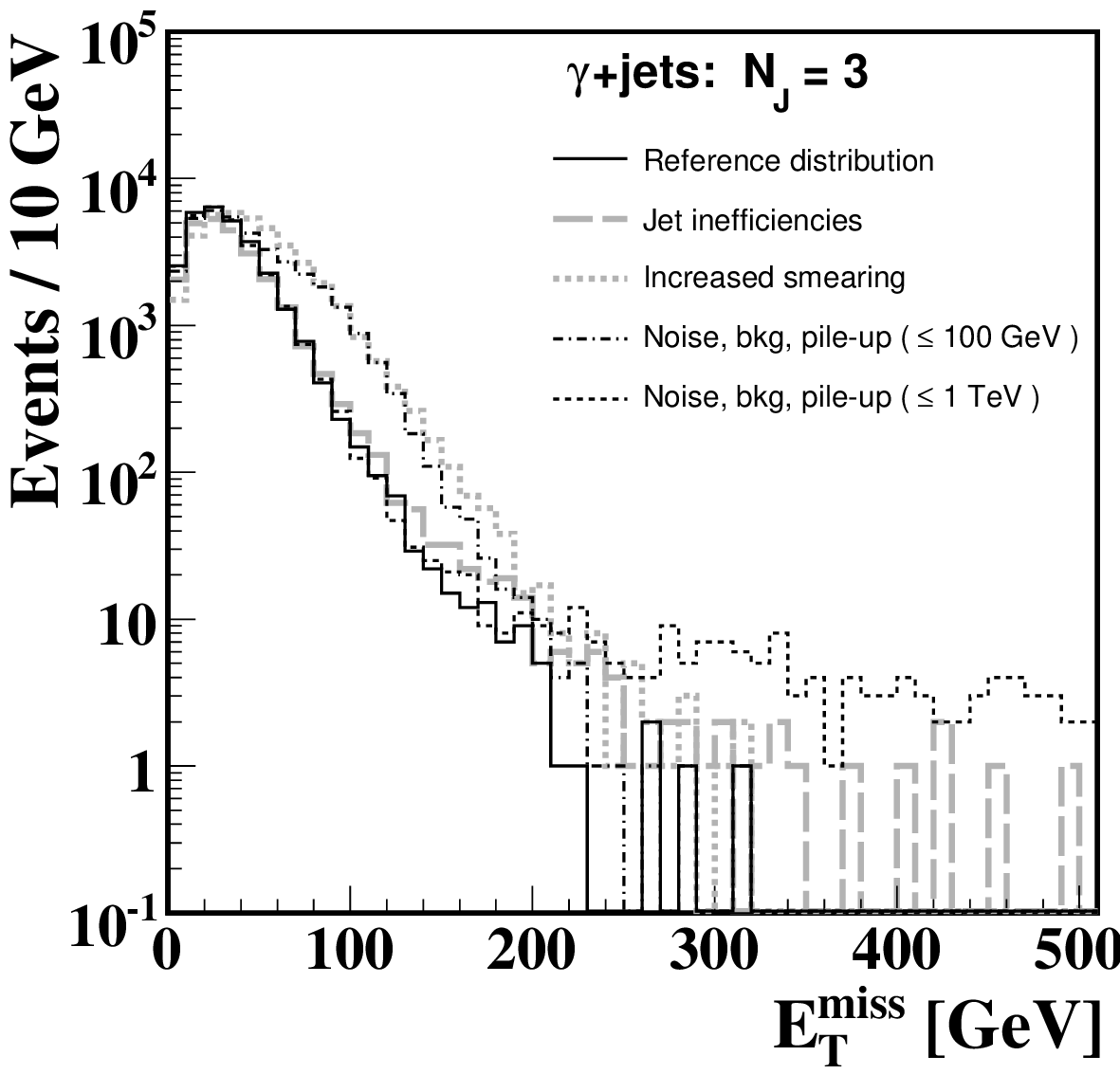,width=3.5in}
\end{center}
\caption{ Illustration of effects associated with jet mis-reconstruction 
          on artificial \met~ in \gammajets~ in the $N_J = 3$ bin for
          the jet \pt~ threshold of 50~GeV. The black line is 
          a reference \met~ distribution from Figure~\ref{figure-3}. 
          Jet reconstruction inefficiencies~(dashed grey),
          increased jet energy smearing~(dotted grey) and
          extraneous energy (dot-dashed black and dotted black)
          from the tests in section~\ref{robustness} and~\ref{limitations}
          significantly increase artificial \met.
          }
\label{figure-6}
\end{figure}

\begin{figure}[ht!]
\begin{center}
\epsfig{file=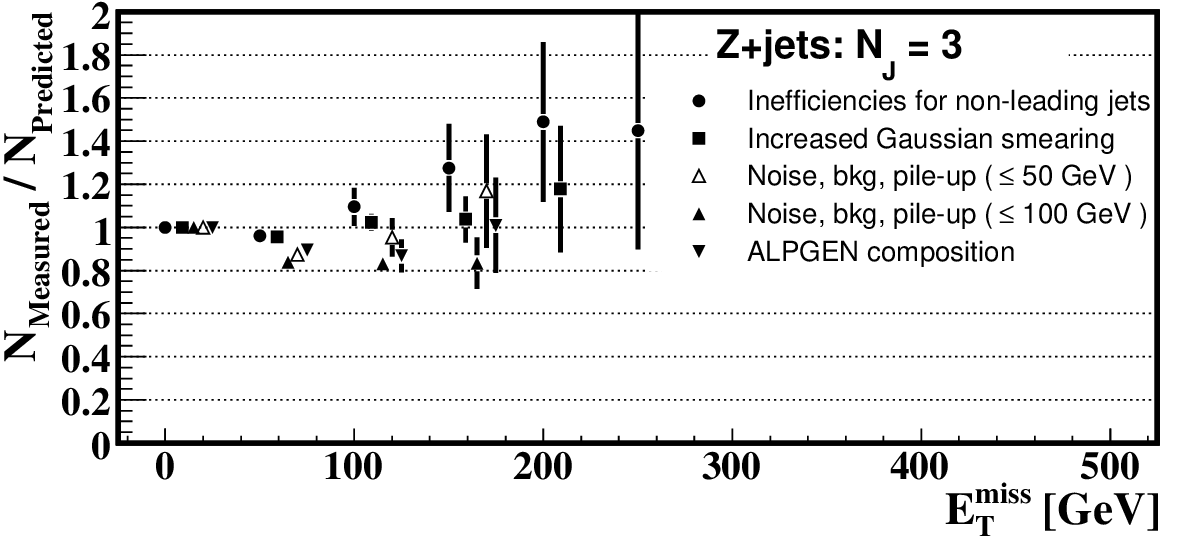,width=3.5in}\\
\epsfig{file=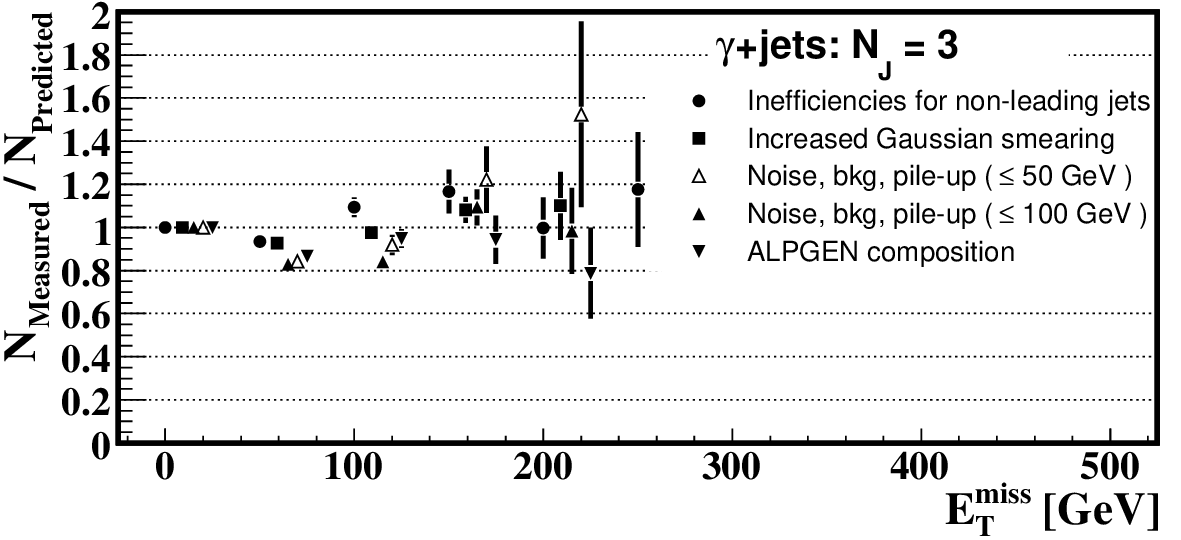,width=3.5in}\\
\epsfig{file=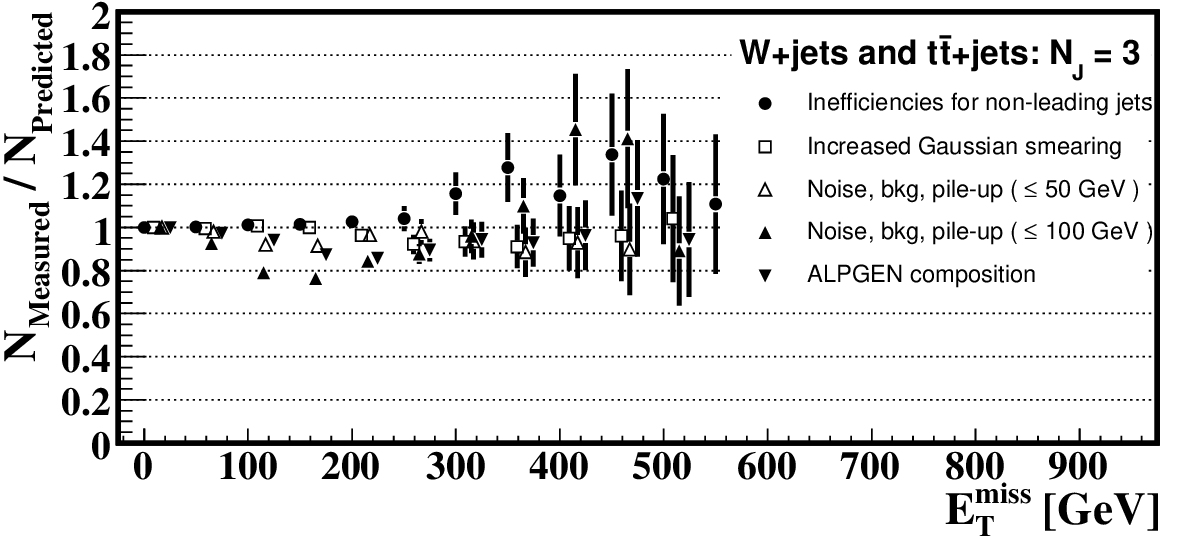,width=3.5in}
\end{center}
\caption{ Ratios of observed and estimated integrated yields
          in \zjets~(top), \gammajets~(middle),  
          \wjets~ and \ttbarjets~(bottom) 
          all for $N_J = 3$ and the 50~GeV jet \pt~ threshold 
          for $N_J$ from robustness tests in section~\ref{robustness}. 
          Circles, squares, triangles-up, triangles-down are for tests with 
          increased inefficiencies for non-leading jets, 
          increased Gaussian jet energy smearing,
          extraneous energy contributions and a modified $N_J$ composition of the QCD sample.
          Note, ratios in the plots are correlated as tests are 
          made using events drawn from the same mock data samples, 
          and yields are integrated upwards. 
	} 
\label{figure-7}
\end{figure}

The goal of this method is to capture effects generating high 
artificial \met~ in-situ using multi-jet QCD events. 
To demonstrate how well the method works, I present a set of tests in 
which increased jet mis-reconstruction 
is introduced. In each test, an identical change to the mock data 
samples for $V$+jets and the QCD sample is made and the analysis procedure is repeated.
Figure~\ref{figure-6} shows how drastic the effect of these changes 
on the \met~ distribution in \gammajets~ can be. 
(The \gammajets~ channel in Figure~\ref{figure-6} is used  
since there is no genuine \met~ in this final state 
and it has a larger yield than \zjets.) 
Test results are presented in Figure~\ref{figure-7} for $N_J = 3$  as 
ratios of observed and estimated integrated yields. For brevity, 
test results for $N_J = 2$~and~$\ge 4$ are not shown but discussed 
and compared to those in Figure~\ref{figure-7}.

Jet reconstruction efficiencies are not equal to unity. 
To test if the method models effects due to undetected jets accurately, 
an identical source of jet inefficiency is introduced in $V+$jets and 
QCD events. I remove jets that fall in a veto cone of 
$\Delta R \equiv \sqrt{ \Delta \eta^2 + \Delta \phi^2 } < 0.8$ 
at $(\eta, \phi) = (0.0, 0.0)$~\cite{jet-loss}, 
where $\phi$ is the azimuthal angle. 
Since softer jets are more likely to be lost, only non-leading jets 
are removed in the veto cone in this test. The effect of this 
inefficiency on the \met~ distribution in \gammajets~ for 
$N_J=3$ can be assessed by comparing the solid line with 
the dashed grey line in Figure~\ref{figure-6}. 
Test results for \zjets, \gammajets~ and \ljetsmet, \wjets~ and \ttbarjets, 
with $N_J=3$ are shown in Figure~\ref{figure-7} in circular markers. 
The increased artificial \met~ tail due to lost jets is modeled 
accurately by the method. The same conclusion holds for $N_J=2$ and~$\ge 4$.

In the next test, the jet energy smearing is increased. 
Two tests are made: (a)~the Gaussian $\sigma(|\vec{p}_T|)$ is 
doubled with the area and shape of the non-Gaussian tails unchanged,  
and (b) the area of the low and high-side non-Gaussian tails 
is doubled with the Gaussian component kept unchanged.
The effect of the additional jet energy smearing in test~(a) on the \met~ 
distribution in \gammajets~ for $N_J=3$ is shows in Figure~\ref{figure-6} 
in the dotted grey line. Ratios of observed and estimated yields 
from test~(a) are in square  markers in Figure~\ref{figure-7} for $N_J=3$. 
Similar results are observed in the other two $N_J$ bins, and in test~(b).
Again, templates constructed from multi-jet QCD events capture effects
from additional jet smearing in-situ so that the level of background at 
high \met~ in $V+$jets is predicted accurately.

Hot cells or noise in the calorimeters, backgrounds from 
the proton beams, cosmic rays, underlying event or additional 
$pp$ interactions in the same bunch crossing contribute extra energy 
and jets erroneously attributed to those produced in $V$+jets 
and QCD processes. 
Since additional jets have higher probability to be soft, 
I test the method's ability to model such effects by adding
extra jets with a soft uniform \pt~ spectrum from 0 to 50~GeV 
with a 20\% probability to each $V$+jets or QCD event.  
These extra jets change $J_T$ and \met, but do not change $N_J$. 
The predictions are good in this test as seen in Figure~\ref{figure-7}
in open triangular~(up) markers.

I repeat the previous test with a uniform \pt~ spectrum of 
additional energy contributions covering the range from 0 to 100~GeV 
added with a 10\% probability to \vjets~ and QCD events. 
This produces a strong effect on the \met~ distribution 
shown for \gammajets~ with $N_J = 3$ in the dot-dashed line 
in Figure~\ref{figure-6}. 
Ratios of observed and estimated yields for the three 
\vjets~ processes in the $N_J=3$ bin are in solid triangular~(up) 
markers in Figure~\ref{figure-7}. 
I find that the prediction is consistent with the measurement
to about 20\% or better in the $N_J = 3$ and $\ge 4$~bins. 
In the $N_J = 2$ bin in \zjets~ and \gammajets, a bias is observed.
The origin of this bias stems from differences in $N_J$ and 
$J_T$ distributions between  $V$+jets and QCD. I discuss it  
and more stringent tests with extraneous energy contributions 
in the next section.

The cross section ratios for $V$+jets and QCD processes: 
$\sigma^{V+{\rm jets}}(n~{\rm jets})$/ $\sigma^{\rm QCD}(n~{\rm jets})$, 
$\sigma^{V+{\rm jets}}(n~{\rm jets})$/ $\sigma^{V+{\rm jets}}(n+1~{\rm jets})$ or
$\sigma^{\rm QCD}(n~{\rm jets})$/ $\sigma^{\rm QCD}(n+1~{\rm jets})$, 
where $n$ is equal to 2 or more, in LHC data are likely to differ from 
that of ALPGEN used in this study. There may also be differences in other 
differential distributions in the jet system of $V$+jets or multi-jet QCD 
events between LHC data and ALPGEN. 
To test how sensitive the method is to such differences, I vary the ALPGEN ratios  
$\sigma^{\rm QCD}(n~{\rm jets})$/$\sigma^{\rm QCD}(n+1~{\rm jets})$, $n\ge2$, 
by a factor of $1.5$ up or down. 
Test results with reduced ratios for $N_J=3$ are shown 
in Figure~\ref{figure-7} in triangular~(down) markers. 
The \met~ predictions are good in this test because they are made on 
an event-by-event basis using QCD events with the same $N_J$ and $J_T$. 
QCD events with other values of $N_J$ and $J_T$ are included only if 
they are misreconstructed, which is a second order effect,  
but it can become significant in regimes where distributions fall 
or rise steeply. Test results for $N_J  = 3$ and~$\ge 4$ are all good.  
For $N_J=2$ in \zjets~ and \gammajets, when the  
$\sigma^{\rm QCD}(n~{\rm jets})$/$\sigma^{\rm QCD}(n+1~{\rm jets})$
ratios  are reduced the prediction improves; 
when the ratios are increased the prediction becomes biased. 
The origin of this bias is the same as in the previous test
and is discussed in the next section.

In conclusion of this section, the quality of the \met~ prediction 
improves at larger $N_J$.
The \met~ prediction is robust for $N_J=3$ and~$\ge 4$ in all tests. 
Events with $N_J=2$ are more susceptible to biases for two reasons. 
First, there are significant differences in the differential 
distributions describing jets in QCD and \vjets:
in QCD di-jets, the jets come mainly from 
leading order parton interactions, while in \vjets, 
the jets are from higher order processes. 
Second, the averaging effects discussed in section~\ref{algorithm} 
are not as strong when the number 
of jets is small. Nevertheless, only two tests for $N_J=2$
are biased in this section. Any other effect that generates 
artificial \met~ in the same manner in the jet system of $V$+jets 
and multi-jet QCD events should be modeled in-situ by the method. 
I next discuss the method's limitations revealed 
in more stringent tests.

\section{Limitations}

\label{limitations}

I increase the degree of jet mis-reconstruction up to a 
point where the method becomes biased to explore the 
boundaries of the domain where the method works.
This allows to understand in greater detail mechanisms 
that may lead to a bias. At the end of this section, 
I discuss how to avoid regimes where the method is biased.

\begin{figure}[h!]
\begin{center}
\epsfig{file=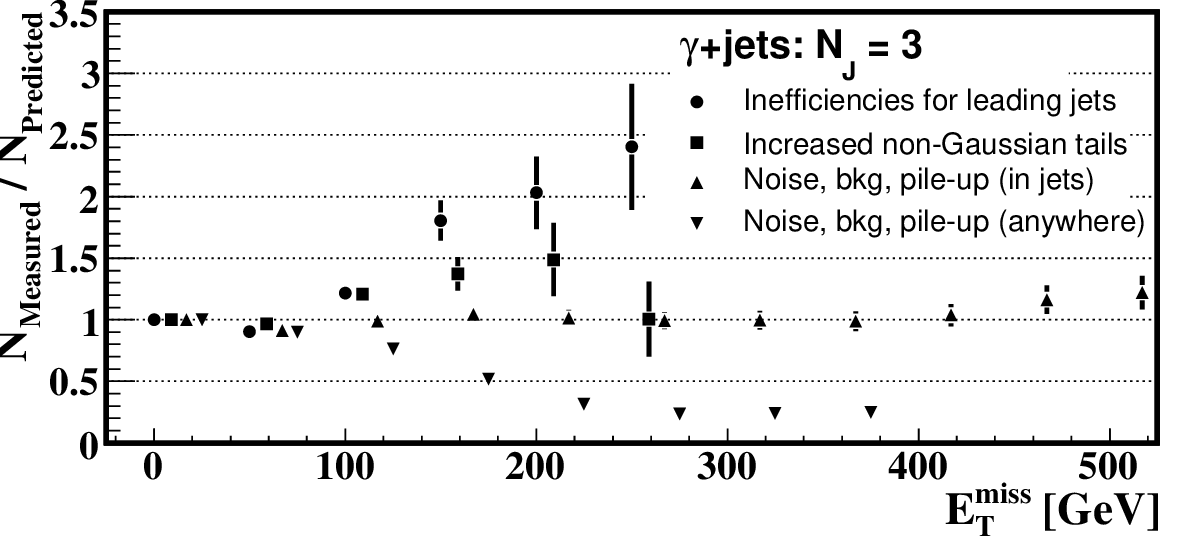,width=3.5in}
\end{center}
\caption{ Ratios of observed and estimated integrated yields
          in \gammajets~ for $N_J = 3$ and the 50~GeV jet \pt~ threshold  
          for $N_J$ from tests in section~\ref{limitations}.
	  Circles, squares, triangles-up and triangles-down are for tests with 
          increased inefficiencies for leading jets, 
          increased non-Gaussian jet energy fluctuations and  
          extraneous energy contributing in jets or anywhere 
          in the calorimeters.
          Note, these ratios are correlated as tests are 
          made using events drawn from the same mock data samples, 
          and yields are integrated upwards. } 
\label{figure-8}
\end{figure}

The test with the jet veto cone introduced in the previous section 
is repeated with a modification such that leading jets  
falling into the veto cone are removed. 
This is a stringent test since leading jets are less likely to 
be undetected. Test results are shown in Figure~\ref{figure-8} 
in circular markers, for brevity, only for \gammajets~ in the $N_J =3$ bin. 
While the prediction partly takes into account the effect of  
undetected leading jets, it underestimates the background 
at high \met~ in that $N_J$ bin in \zjets~ and \gammajets.
The prediction is biased because in QCD events \met~ is always 
less than $J_T$, by the definition of  \met~ in section~\ref{experiment}. 
In \vjets, $V$ is a $Z$ or $\gamma$ here, \met~ can be greater 
than $J_T$ when the leading jet recoiling against an energetic 
$V$~boson in the transverse plane is lost. 
The $V$+jets events with \met~ larger than $J_T$ can not 
be modeled by the algorithm in section~\ref{algorithm}.
This bias is larger for $N_J = 2$, while in the $N_J \ge 4$ bin, 
the prediction is good for both \zjets~ and \gammajets.
In \ljetsmet, \wjets~ and \ttbarjets~ combined, 
due to a genuine \met~ contribution from neutrinos 
to the full \met, this bias does not appear.

I repeat the test with increased jet energy mis-measurements
after tripling the area of the lower non-Gaussian tail 
in the jet response function and reducing the magnitude 
of its slope on the logarithmic scale of the lower plot 
in Figure~\ref{figure-1} by a factor of 2.  
The prediction is biased in the $N_J = 2$ bin for both 
\zjets~ and \gammajets. The quality of the prediction improves 
in the $N_J = 3$ bin, shown in Figure~\ref{figure-8} in square
markers for \gammajets, 
and it is good for $N_J \ge 4$ in \zjets~ and \gammajets.
One should expect a bias for large low-side tails in the jet response 
function appearing 
via the same mechanism operating in the previous test. 
The $V$+jets events containing jets fluctuated down in \pt~ can have $J_T$ 
that is less than \met. 
Such events can not be modeled by the algorithm of section~\ref{algorithm}.
In the \ljetsmet~ final state, the prediction of the full \met, 
which includes the neutrino momentum, is good in all $N_J$ bins.
Note, for a large low-side tail in the jet response function, 
the jet energy scale may become biased. 
Effects due to a jet energy scale offset 
are discussed in section~\ref{jet-energy-scale}.

I make two tests with a harder spectrum of additional energy contributions
unrelated to \gammajets~ and QCD events. 
In these tests, the spectrum of additional energy is uniform in 
\et~ from 0 to 1~TeV added with a 1\% probability to both \vjets~ and QCD. 
Since there is no genuine \met~ in \gammajets~ and QCD, 
the requirement on 
$| \Delta\phi^{\rm lead\; jet-{\it E}_T\hspace{-3.4mm}/\hspace{1.5mm}} |$~(section~\ref{experiment}) 
that \metvector~ and the leading jet be not 
aligned in the transverse plane removes a fraction of events 
with high $E_T$ extraneous contributions.
In the first test, additional energy depositions contribute only to jets 
that are above the \pt~ threshold for $N_J$. 
The \met~ distribution in \gammajets, $N_J = 3$, is shown 
in the dotted black line in Figure~\ref{figure-6} with a large 
artificial high \met~ tail. Ratios of observed and estimated 
yields are in triangular~(up) markers in Figure~\ref{figure-8}. 
In \gammajets, the prediction is good for $N_J = 3$ and~$\ge 4$,
and it is biased in the $N_J=2$ bin for the following reason. 
The $J_T$ spectrum in QCD events tends to be softer than that 
in \vjets~ events with the same $N_J$. 
(This effect is most pronounced for $N_J=2$.)
The fraction of soft QCD  multi-jet events promoted to higher 
$J_T$ by extraneous energy depositions tends to be larger 
than that fraction in $V$+jets. 
Since such events have larger \met~ due to the extraneous 
energy depositions unbalanced in the transverse plane, the 
level of background at high \met~ is overestimated.

In the second test, extraneous energy contributions are added 
randomly in the $\eta-\phi$ plane so that $N_J$ also tends to increase. 
Ratios of observed and estimated yields for \gammajets, $N_J = 3$, 
are in triangular~(down) markers in Figure~\ref{figure-8}. 
The prediction overestimates the background in all $N_J$ bins. 
This happens because 
$\sigma_{\rm QCD}(n{\rm \; jets}) / \sigma_{\rm QCD}(n+1{\rm \; jets})$, $n\ge2$,
is higher than
$\sigma_{V+{\rm jets}}(n{\rm \; jets}) / \sigma_{V+{\rm jets}}(n+1{\rm \; jets})$
in the mock data samples so that the fraction of events 
with $N_J = n$ reconstructed erroneously in the  $N_J = (n+1)$ bin 
due to an extra energy deposition is higher in QCD compared to 
that fraction in $V+$jets. (Again, this effect is most pronounced for $N_J=2$.)
Since these mis-reconstructed events have larger \met, 
the prediction overestimates the background.  
The mechanisms leading to a bias described in this and the previous 
paragraphs are also responsible for biases noted in the previous sections.

Test results with extraneous energy contributions for \zjets~ 
are qualitatively similar to those for \gammajets. In \ljetsmet, 
the biasing effects discussed above are intertwined with additional 
effects due to the presence of a neutrino in the final state and 
the \ttbarjets~ contribution. The genuine \met~ from the neutrino 
makes the requirement on 
$| \Delta\phi^{\rm lead\; jet-{\it E}_T\hspace{-3.4mm}/\hspace{1.5mm}} |$ 
less efficient in suppressing high $E_T$ extraneous contributions. 
The \ttbarjets~ events contribute to further differences in $N_J$ and $J_T$ 
spectra between \wjets~ and QCD. 
I find that in \ljetsmet~ the prediction tends to overestimate the 
background in the tests with extraneous energy depositions, 
and the quality of the prediction improves with $N_J$.

Regimes with severely misreconstructed events where 
the method may become biased need to be avoided.
By imposing event quality criteria or improving the jet 
reconstruction, $e.g.,$ using the tracking systems,~\cite{jets-review} 
one can reduce the number of such events.
Moderately mis-reconstructed events are modeled in-situ by the method.
The \vjets~ sample with $N_J=2$ is the most challenging for this method.
This makes two jet events a good sample with which to validate the 
algorithm in data.
The method performs better at higher $J_T$ and $N_J$, 
where the sensitivity to new physics is higher. 
There are several reasons for that:
a)~there are fewer differences between the hadronic systems 
in $V$+jets and QCD, b)~the averaging effects over $V$+jets 
and QCD events are stronger and  
c)~the jet reconstruction performs better at higher jet \pt.

\section{ \ttbarjets }

\label{ttbar-section}

\begin{figure}[h!]
\begin{center}
\epsfig{file=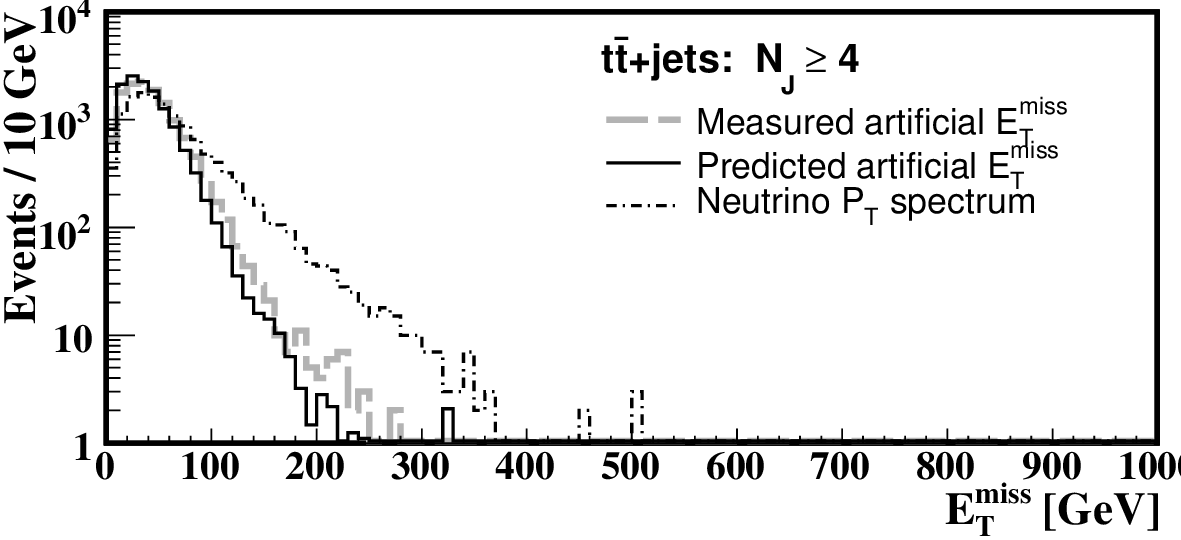,width=3.5in} \\
\epsfig{file=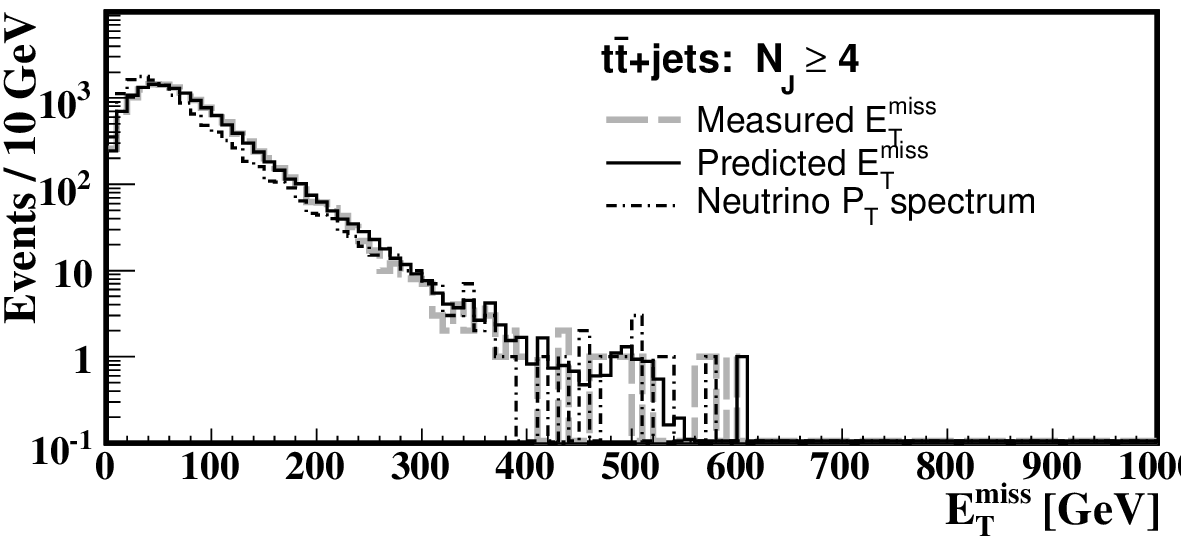,width=3.5in}
\end{center}
\caption{ Top: artificial \met~ in \ttbarjets~(dashed) and 
               its prediction (solid) for $N_J \ge 4$ and 
               the 50~GeV jet \pt~ threshold for $N_J$.  
          Bottom: full \met~ (dashed) and its prediction (solid) 
               combining both the neutrino \pt~ spectrum and the artificial
               \met~ prediction for the same $N_J$ and jet \pt~ selection. 
               In both plots the neutrino spectrum is shown
               as the dot-dashed line. 
               }
\label{figure-9}
\end{figure}

SM \ttbarjets~ events, where $t\bar{t} \rightarrow l\nu_lb\bar{b}q\bar{q}$,
constitute a dominant background in the \ljetsmet~ signature for $N_J \ge 3$. 
The shapes of $N_J$ and $J_T$ spectra in these events differ from those in 
QCD events collected for templates and from those in $V$+jets.
The calorimeter response to $b$-jets in \ttbarjets~ differs
from that of light quark and gluon jets~\cite{b-jets}.
These effects lead to a bias in the prediction of artificial \met~ in \ttbarjets. 
To demonstrate this bias clearly, Figure~\ref{figure-9}~(top) shows the 
artificial \met~ in \ttbarjets~ for $N_J \ge 4$ (dashed line), where the neutrino 
four-momentum is assumed to be measured so that it is included in the \met~ 
calculation,  and its prediction using QCD templates~(solid line). 
Note, at large \met, this bias is an order of magnitude smaller compared 
to the genuine \met~ from the neutrino in the final state having
the \pt~ spectrum shown in the dot-dashed line in the same Figure.  
When the neutrino \pt~ spectrum is combined with the 
artificial \met~ in the full \met~ prediction, the bias becomes  
insignificant as seen in Figure~\ref{figure-9}~(bottom).

The artificial \met~ is a dominant contributor in events with 
small genuine \met. 
Figure~\ref{figure-9}~(bottom) shows that 
the accuracy of its prediction is sufficient to model 
the full \met~ distribution at small \met. 
At high \met, the missing momentum from the neutrino dominates over 
artificial \met~ so that the accuracy of the full \met~ prediction 
is highly dependent on how well the neutrino spectrum is modeled.
The modeling of neutrino \pt~ spectra is discussed in section~\ref{nu-spectrum}.

Despite the fact that the bias in the artificial \met~ prediction
for \ttbarjets~ is insignificant in the full \met~ prediction in \ljetsmet, 
it is instructive to examine how it behaves when selection criteria 
or the algorithm of section~\ref{algorithm} are modified. 
Two observations can be made.
First, the bias becomes smaller when the jet \pt~ threshold for \met~ 
and $J_T$ is reduced or the $\eta$~coverage for jets is increased 
since the total energy is collected to a fuller extent with more 
inclusive requirements. Optimal requirements on these variables can only
be determined using data because at smaller \pt~ and larger $|\eta|$
more noise and backgrounds are expected.
Second, in \ttbarjets, there tends to be more jets included 
in the \met~ and $J_T$ calculations that are below the jet \pt~ 
threshold for $N_J$. Since the jet resolution improves as the jet \pt~ grows,
the prediction can be improved by making \met~ templates in coarse 
bins of $R(J_T) = J_T^{\rm high}/ J_T$, where $J_T^{\rm high}$ 
is a scalar sum of jet \pt's for jets above the \pt~ threshold 
for $N_J$.  Alternatively, the same effect can be achieved 
by modifying the composition of the QCD sample used for templates.
Finally, the modeling of $b$-jets  in \ttbarjets~ can be improved 
by removing a fraction of jet \pt~ measurements in QCD events 
that is expected to be carried by muons and neutrinos in 
semileptonic decays of beauty and charm quarks in $b$-jets.

\section{Jet energy scale}

\label{jet-energy-scale}

Jet energy measurements could be systematically biased in 
early data. 
Let us consider a case when jet energies are under-measured 
uniformly in jet \pt. 
Such mis-measurements cancel to first order in \met~ measurements in QCD events.
In $V$+jets, since the jet system recoils against the $V$, 
the jet energy mis-measurements add up coherently along 
the $V$ direction in the transverse plane. 
To avoid a bias due to this difference, the jet energy scale needs 
to be calibrated. Since the method is capable to model large tails 
in the jet response function, a precise calibration of the jet 
energy scale as a function of $\eta$ and $\phi$~(azimuthal angle) 
is not required. The jet energy scale can be calibrated with 
sufficient accuracy using standard techniques based on \gammajets~ 
and \zjets~($N_J \le 2$) processes~\cite{jets-review} in very early data. 

\begin{figure}[h!]
\begin{center}
\epsfig{file=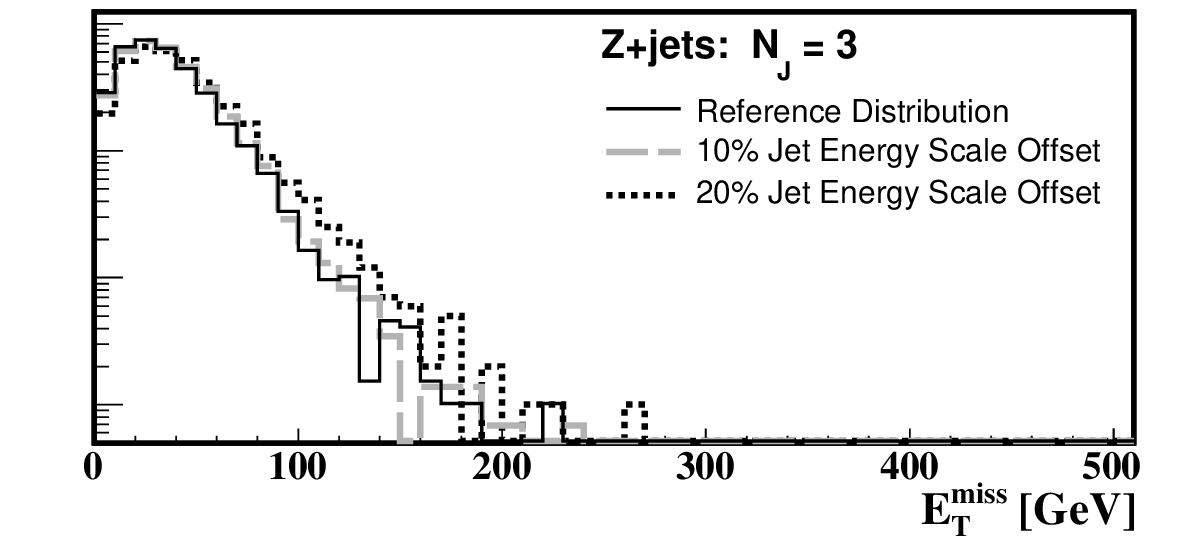,width=3.5in}\\
\epsfig{file=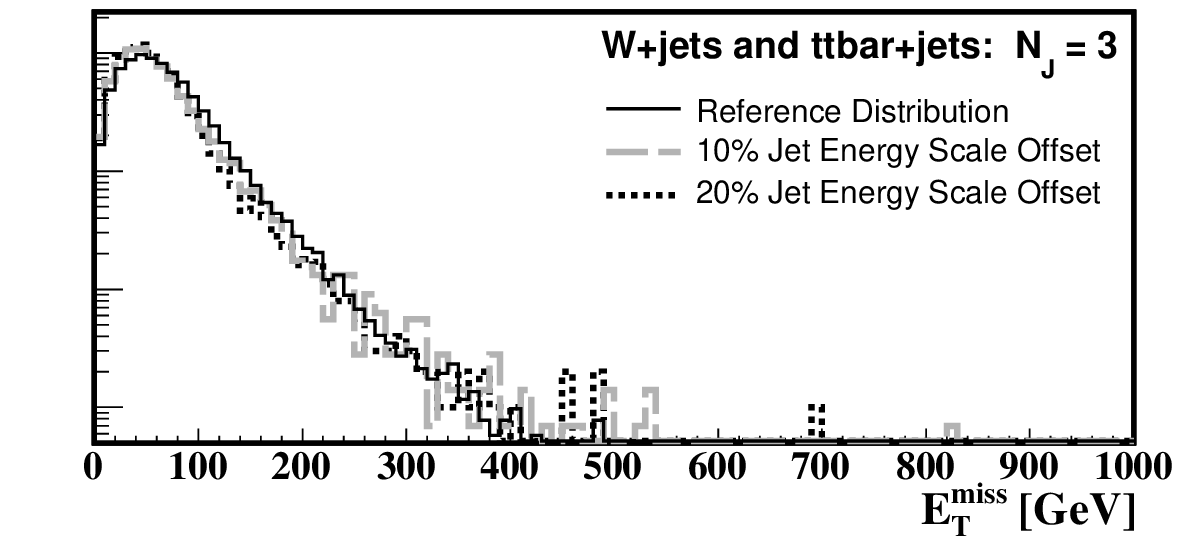,width=3.5in}
\end{center}		
\caption{ The \met~ distributions without~(solid) and with 20\%~(dotted) 
          or 10\%~(dashed) jet energy scale offset in \zjets~(top) and 
         \ljetsmet~(bottom) for $N_J = 3$ with the 50~GeV jet threshold for $N_J$. 
         The distributions in each plot are normalized to the same area. }
\label{figure-10}
\end{figure}

Figure~\ref{figure-10} gives a comparison of \met~ distributions 
in \zjets~(top)  and \ljetsmet~(bottom) for $N_J = 3$ without~(solid) 
and with a 10\%~(dashed) and 20\%~(dotted) jet energy scale offset 
downwards uniform in jet \pt.  
The prediction becomes good for a 10\% or smaller offset in \zjets~ 
and \gammajets.  
One may reduce the effect from a residual jet energy 
scale offset on \met~ in \zjets~ and \gammajets~ by normalizing 
the predicted \met~ shape to the observed distribution in 
the small \met~ region, for example, for \met~ $\in [50,100]~{\rm GeV}$. 
Demands on the precision of the jet energy calibration in \ljetsmet~
are higher. 
Finally, even before the jet energy scale is calibrated, 
one can make a search in the projection of \metvector~ on the axis 
perpendicular to the $V$~direction~(the $l$~direction in \ljetsmet) 
in the transverse plane, $E_{TT}\hspace{-6.1mm}/\hspace{4.0mm}$.
Searches in $E_{TT}\hspace{-6.1mm}/\hspace{4.0mm}$
are less sensitive to effects associated with 
the jet energy scale offset since those lead to a bias 
along the $V$~direction.

\section{ Neutrino spectra in $W$ decays }

\label{corrections}
\label{nu-spectrum}

In the $l+{\rm jets}+$\met~ signature, dominated by \wjets~ and 
\ttbarjets, there are one or more undetected neutrinos in the final state. 
To model \met~ in these events, one needs a prediction
or a measurement of the neutrino \pt~ spectra, which can be
combined with \met~ resolution predictions from QCD templates. 
The neutrino \pt~ spectra could be obtained from MC simulation.
Or, the neutrino \pt~ spectra can be modeled in a data-driven 
manner using charged lepton \pt~ spectra as described in this section. 

\subsection{ $W \rightarrow l  \nu_l$ }

\label{corrections-munu}

\begin{figure}[h!]
\begin{center}
\epsfig{file=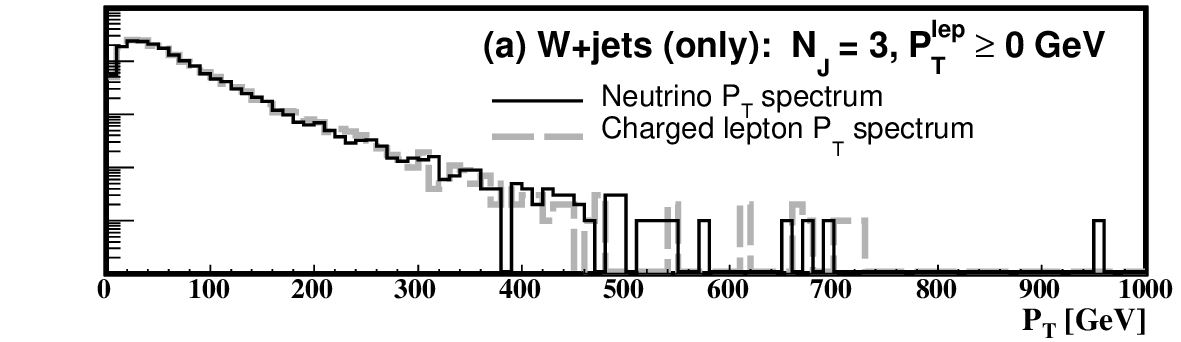,width=3.5in}\\
\epsfig{file=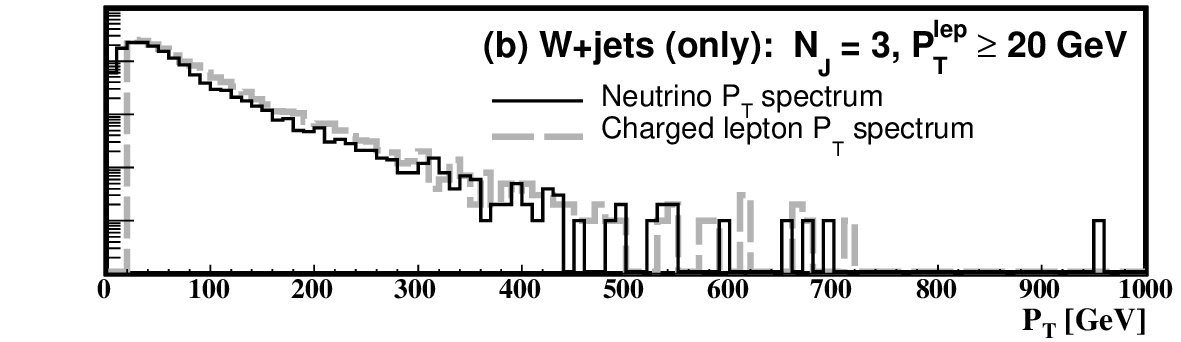,width=3.5in}\\
\epsfig{file=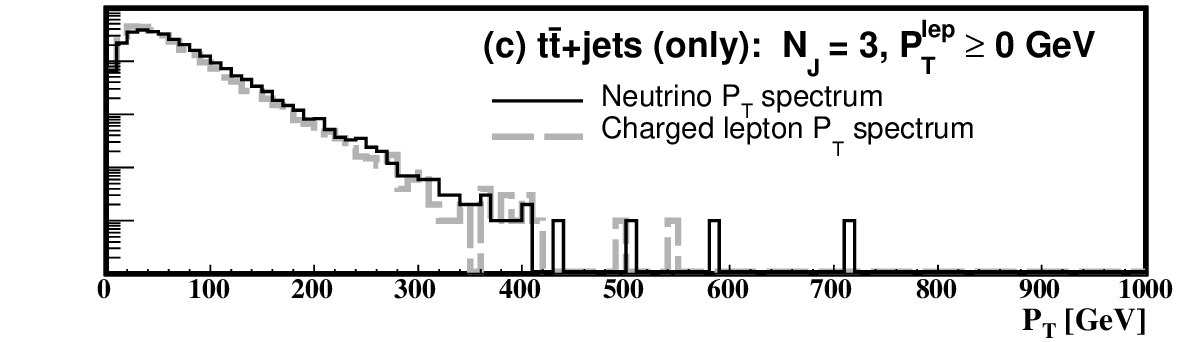,width=3.5in}\\
\epsfig{file=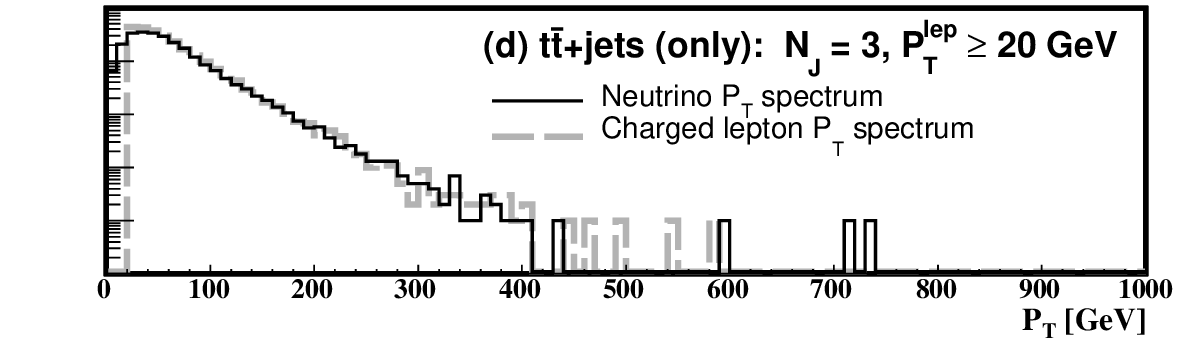,width=3.5in}\\ 
\epsfig{file=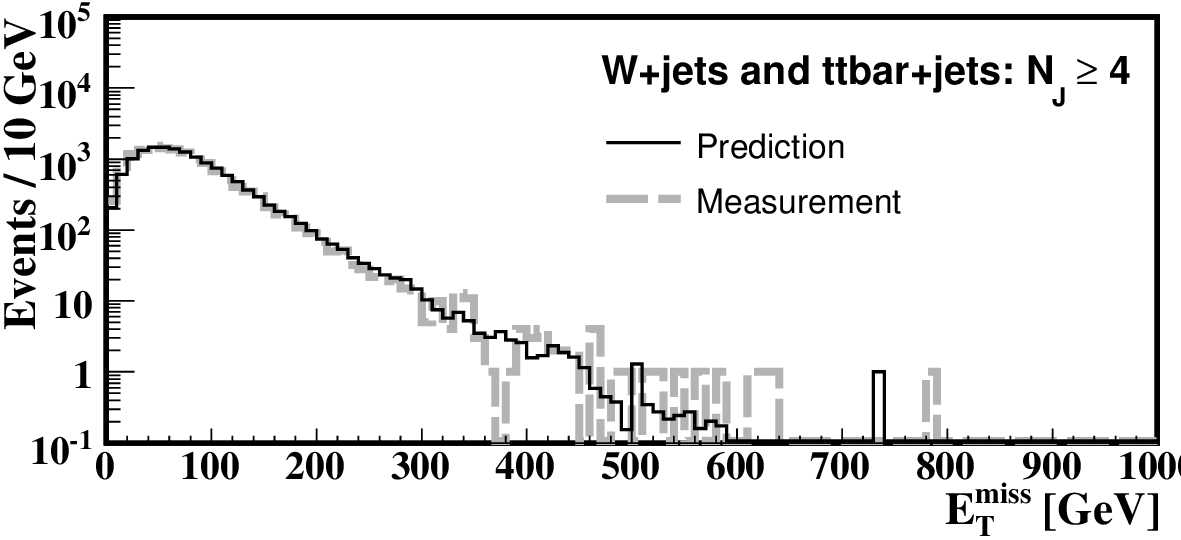,width=3.5in}
\end{center}
\caption{ Comparisons of charged lepton and neutrino spectra in \wjets~(a,b) 
	  and \ttbarjets~(c,d) without~(a,c) and with~(b,d) charged lepton
          \pt~ thresholds all for $N_J = 3$. The lower plot shows a \met~ prediction
          based on the charged lepton spectrum in \ljetsmet, $N_J \ge 4 $, 
          without any corrections. 
          In all plots, the threshold for $N_J$ is 50~GeV. }
\label{figure-11}
\end{figure}

The solid and dashed lines in plot~(a) of Figure~\ref{figure-11} 
are the neutrino and charged lepton \pt~ spectra in \wjets~ for 
$N_J =3$, $W \rightarrow l \nu_l$, passing all selection of 
section~\ref{experiment} but the requirement on 
the charged lepton \pt~ of at least 20~GeV.
It is seen that the two \pt~ spectra have consistent shapes so 
that the charged lepton \pt~ spectra can be used to model 
the neutrino \pt~ spectra in \wjets.
Note, the $W$~bosons in $W+$jets tend to be produced in the 
transverse-minus helicity state~(left-handed) rather than in
the transverse-plus helicity state~(right-handed). 
These polarization effects are present even in the transverse plane
for $N_J \ge 2$ so that $W^+$~($W^-$)~bosons in \wjets~ tend to 
produce charged leptons with a \pt~ spectrum that is softer~(harder) 
compared to the neutrino \pt~ spectrum. 
However, in the entire \wjets~ sample, 
$W^+$ and $W^-$ combined, these polarization effects are averaged 
and largely disappear~\cite{w-polarn} so that the charged lepton and neutrino 
\pt~ spectra have very similar shapes seen in plot~(a) of Figure~\ref{figure-11}.
The application of a \pt~ threshold on the charged lepton makes 
its spectrum harder, while the neutrino spectrum becomes softer, 
as seen in plot~(b) of the same Figure.
To model the neutrino spectrum using the reconstructed charged 
lepton spectrum in \wjets, the effect of the charged lepton \pt~ 
threshold and the residual differences between charged lepton 
and neutrino \pt~ spectra need to be corrected. The corrections 
can be obtained from MC simulation.

The solid and dashed lines in plot~(c) of Figure~\ref{figure-11} 
are the neutrino and charged lepton \pt~ spectra in \ttbarjets~ 
for $N_J=3$, $t\bar{t} \rightarrow l\nu_lb\bar{b}q\bar{q}$, 
after the selection of section~\ref{experiment} but 
without a threshold requirement on the charged lepton \pt. 
In the SM, 30\% of $W^+$~($W^-$)~bosons in $t$~($\bar{t}$)~quark 
decays are left-handed~(right-handed) and the rest are 
longitudinally polarized~\cite{top-polarization}. 
Left-handed $W^+$ and right-handed $W^-$~bosons tend to produce 
charged leptons with a \pt~ spectrum that is softer compared to 
the neutrino \pt~ spectrum as seen in plot~(c).
Since the two spectra have similar shapes, it is possible
to use the charged lepton \pt~ spectrum to model the neutrino 
spectrum \pt~ in \ttbarjets.  
Again, when a charged lepton \pt~ threshold is applied, the 
charged lepton spectrum becomes harder while 
the neutrino spectrum becomes softer, 
which leads to a higher consistency between 
the two spectra seen in plot~(d).
Nevertheless, the effects of the $W$ polarization 
in top decays and the event selection, mainly due 
to the charged lepton \pt~ threshold, in \ttbarjets, 
in general, need to be corrected.

In order to determine corrections to the charged lepton spectra 
for \wjets~ and \ttbarjets~ from MC simulation, one needs to 
measure the shape of the \pt~ dependence of lepton 
reconstruction efficiencies and the relative fractions 
of \wjets~ and \ttbarjets~ in the data sample. The former can be 
readily done via a standard technique based on $Z\rightarrow l^+l^-$ 
decays~\cite{z-and-w-cdf}. The latter should come from 
an independent measurement. With these two ingredients, 
corrections can be determined from MC simulation.

Since corrections to the charged lepton spectra are small, 
the reliance on details of MC simulation to determine 
the neutrino \pt~ spectra is minimal. 
For a 20 or 15~GeV threshold on charged lepton \pt, 
no corrections are required to predict the \met~ distributions 
in \ttbarjets~ in all $N_J$ bins to 20\% or better 
in the mock data samples. Corrections are needed for \wjets.
The lower plot in Figure~\ref{figure-11} shows the \met~ distribution 
and its prediction in \ljetsmet, \wjets~ and \ttbarjets~ combined, 
for $N_J \ge 4$ based on the charged lepton spectrum without corrections. 
Since \ttbarjets~ dominates over \wjets~ in the $N_J\ge4$~($N_J=3$) bin,
the prediction is good to 15\%~(25\%) at high \met~ without corrections.
The $N_J\ge4$ bin, where the prediction is the most robust, 
is likely to have the highest sensitivity to a new physics 
contribution compared to lower jet multiplicity events.

\subsection{ $W \rightarrow \tau \nu_\tau$ }

In the $l$+jets+\met~ signature, there is background 
from tauonic $W$ decays in \wjets~ and \ttbarjets. 
Tauonic $W$ decays produce at least one additional 
neutrino that is a source of differences between 
the charged lepton and neutrino \pt~ spectra.

There are two types of tauonic $W$ decays that contribute
significant background:  
(1) \wjets~ and \ttbarjets, where 
$W^- \rightarrow \tau \bar{\nu}_\tau$ with 
$\tau \rightarrow  l \bar{\nu}_l \nu_\tau$, 
and 
(2) $t\bar{t}$~ events, where $W^- \rightarrow l \bar{\nu}_l$
and $W^+ \rightarrow \bar{\tau} \nu_\tau$ with 
$\bar{\tau} \rightarrow ({\rm hadrons}\; \bar{\nu}_\tau)$.
The contribution from tauonic $W$ decays is an order of magnitude
smaller compared to that from $W \rightarrow l \nu_l$ decays.
(The tauonic background of type~2 can be suppressed 
by vetoing events with isolated single hadronic tracks.)
The $\tau$ branching fractions are well known. Therefore, 
the effects from $W \rightarrow \tau \nu_\tau$ on \met~ predictions 
can be well-modeled by an additional smooth correction to 
the charged lepton \pt~ spectra that can be determined by MC simulation.
Since the contribution from tauonic $W$ decays is smaller compared 
to that from $W \rightarrow l \nu_l$ decays in $l$+jets+\met, 
corrections for them are not discussed further in this paper.

\section{ Systematic uncertainties }

\label{systematics}

Systematic uncertainties need to account for the statistical 
precision and biases of the method's background predictions
at high \met. Mechanisms by which biases may appear are discussed in 
section~\ref{limitations}. The method's susceptibility to them 
can  be studied in both data and MC. 

A sample of events with $N_J = 2$ for a small jet \pt~ 
threshold, such as 50~GeV, is more sensitive to biases.
The relative contribution from new physics can not be 
large in this sample. Therefore, these events can 
be used to validate the method's performance and place 
an upper bound on its biases at higher $N_J$ and jet \pt. 
Similarly, the application of event quality criteria are 
expected to reduce the number of severely misreconstructed 
events that may lead to biases in the prediction of artificial 
\met. By varying the event quality 
selection criteria, one can determine if the method is 
subject to such biases or estimate their size. 
An excess due to a new physics contribution 
should be stable under variations of these criteria.

In the $l$+jets+\met~ channel, lepton \pt~ spectra are used to model 
neutrino \pt~ spectra. There are several sources of 
the systematic uncertainty associated with this modeling as MC is used to 
obtain corrections to the charged lepton \pt~ spectra. 
Because these corrections are small, uncertainties 
due to MC used to extract them enter only at second order. 
They can be estimated by varying the composition of the 
MC samples used to measure them and the reconstruction 
efficiencies of leptons and jets within their uncertainties. 
The uncertainties in the composition of the MC samples 
should come from an independent measurement
of the relative \wjets~ and \ttbarjets~ cross sections 
for different $N_J$. Note, in section~\ref{corrections-munu},
it is demonstrated that these corrections may be 
negligible for $N_J \ge 4$ to obtain a prediction 
at high \met~ to 20\% or better in early data.

The QCD background to signal events with one or more fake 
leptons or photons, cross-feeds among \vjets~ processes and 
other secondary backgrounds are not considered in this paper. 
These backgrounds as well as di-leptons from \ttbarjets~ could 
be accounted in the \met~ distributions and their predictions 
by the following procedure.
For each significant secondary background contribution, 
one can obtain a control sample in data and estimate 
the number of events from this background contribution 
in the entire search sample~\cite{previous-methods}\cite{z-and-w-cdf}. 
Next, one can measure the \met~ distribution and make 
its prediction in that control sample, and normalize both to 
the expected number of events for this background contribution.
The \met~ distribution and its prediction from the control sample 
could then be subtracted from the \met~ distribution and its 
prediction for the entire sample, respectively.

A large new physics contamination to QCD at large $J_T$,
in general, may bias the prediction at large \met~ and 
hide a new physics contribution to \vjets. I find that even
under the most optimistic scenarios for new physics cross sections 
such a contamination does not lead to a significant bias.

Even though the reliance on MC is much reduced in this 
method, MC can be used to validate the method and constrain 
its systematic biases as is done in this paper. 
Nevertheless, a study of control data samples is 
needed to develop, optimize and validate the final 
algorithm and to quantify its systematic uncertainties.

\section{ Predictions with Signal }

\label{prediction-with-signal}

\begin{figure*}[ht!]
\begin{center}
\begin{minipage}{7.1in}
\begin{center}
\epsfig{file=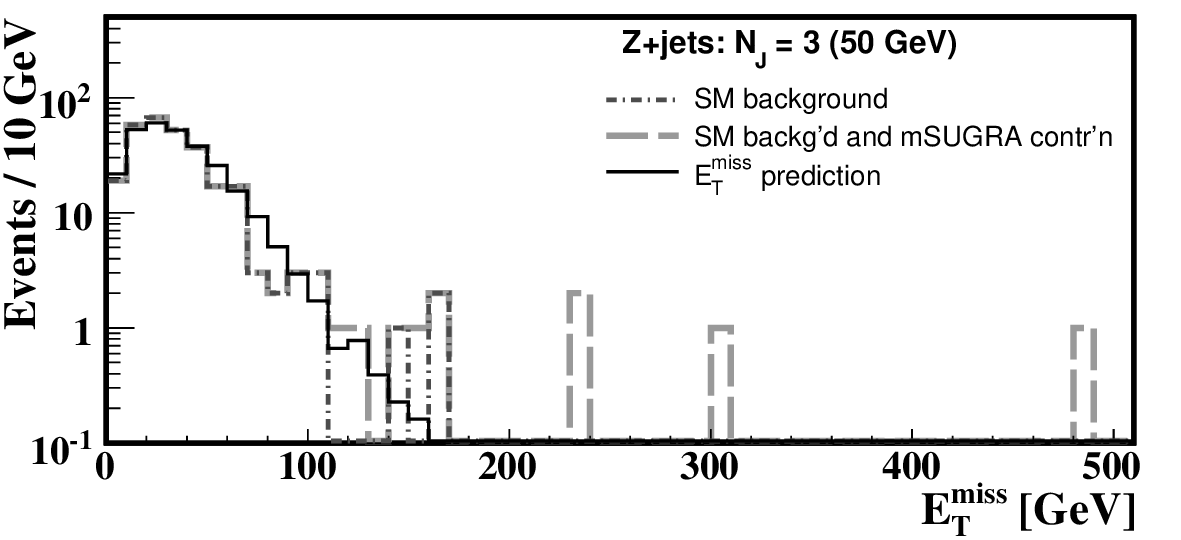,width=3.5in}
\epsfig{file=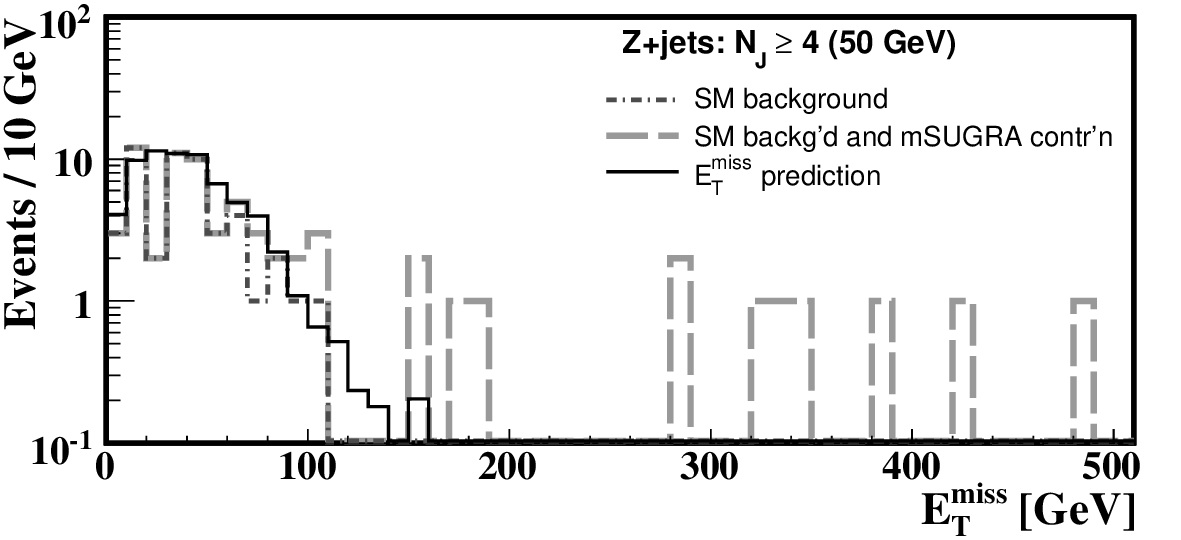,width=3.5in}
\end{center}
\end{minipage}
\begin{minipage}{7.1in}
\begin{center}
\epsfig{file=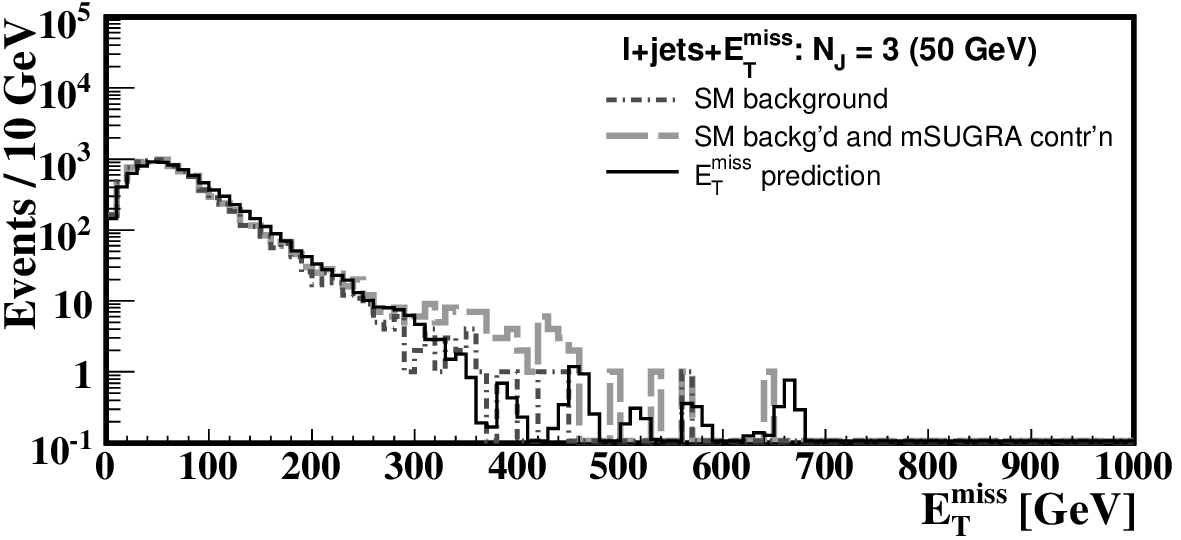,width=3.5in}
\epsfig{file=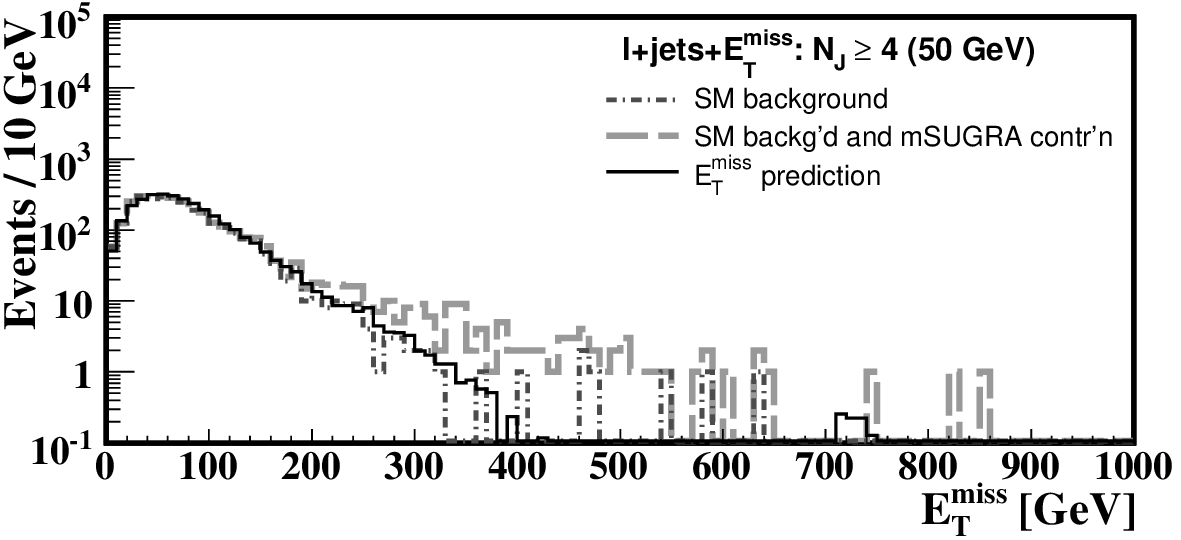,width=3.5in}
\end{center}
\end{minipage}
\end{center}
\caption{ Observed~(dashed) and predicted~(solid) SM \zjets~(top) and \ljetsmet~(bottom) 
          for $N_J = 3$~(left) and $N_J \ge 4$~(right) with new physics contributions 
          from mSUGRA benchmarks~\cite{{cms-msugra}}. 
	  The dot-dashed lines highlight the SM contributions. 
          The \pt~ threshold for $N_J$ is 50~GeV.
          The plots correspond to 200~pb$^{-1}$ at $\sqrt{s} = 14$~TeV.
	}
\label{figure-12} 
\end{figure*}

\begin{figure*}[ht!]
\begin{center}
\begin{minipage}{7.1in}
\begin{center}
\epsfig{file=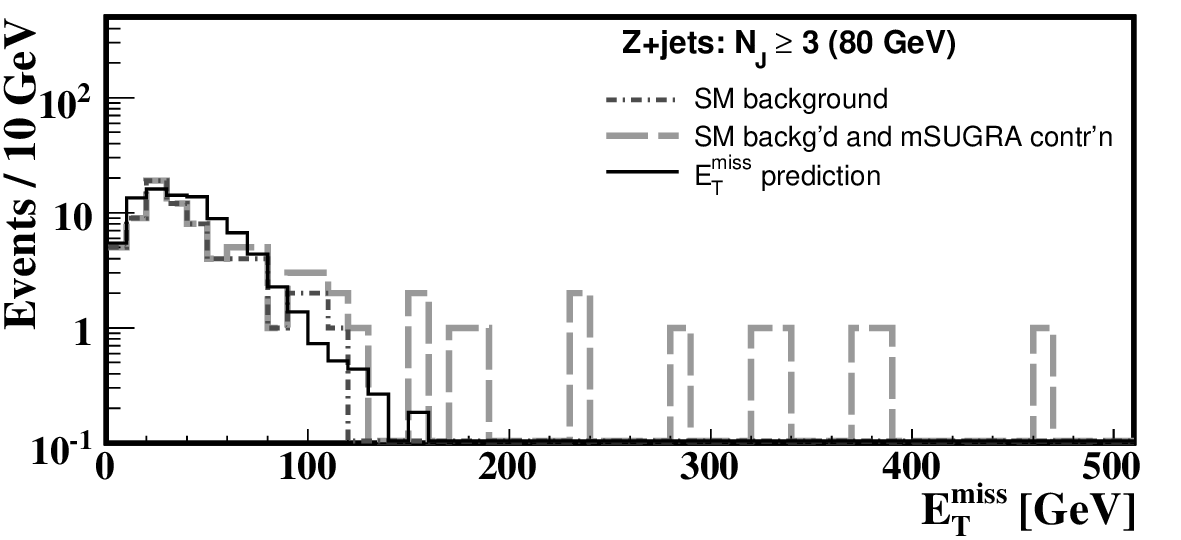,width=3.5in}
\epsfig{file=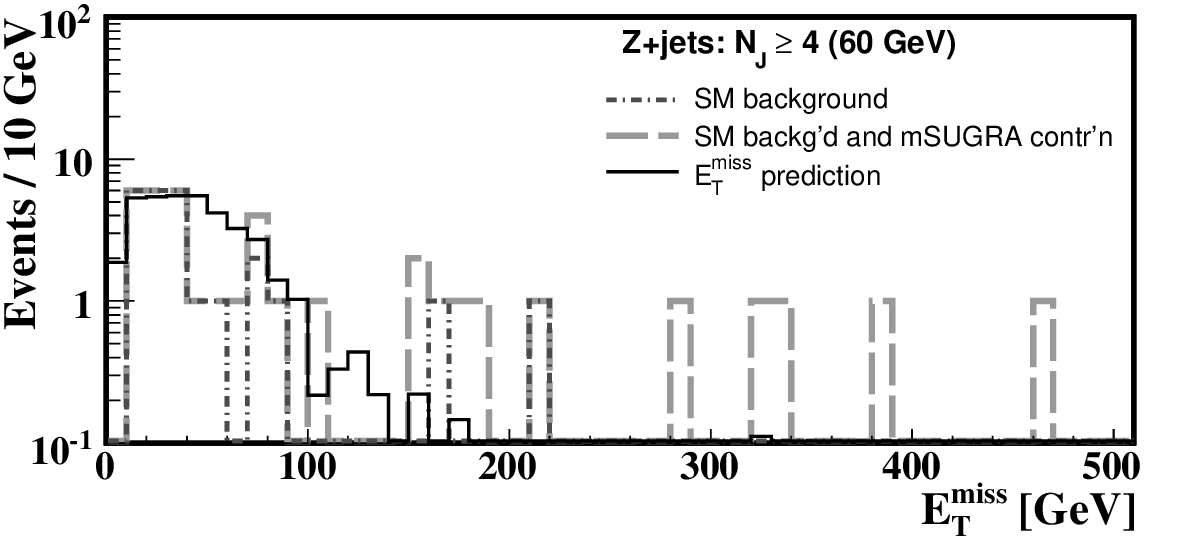,width=3.5in}
\end{center}
\end{minipage}
\begin{minipage}{7.1in}
\begin{center}
\epsfig{file=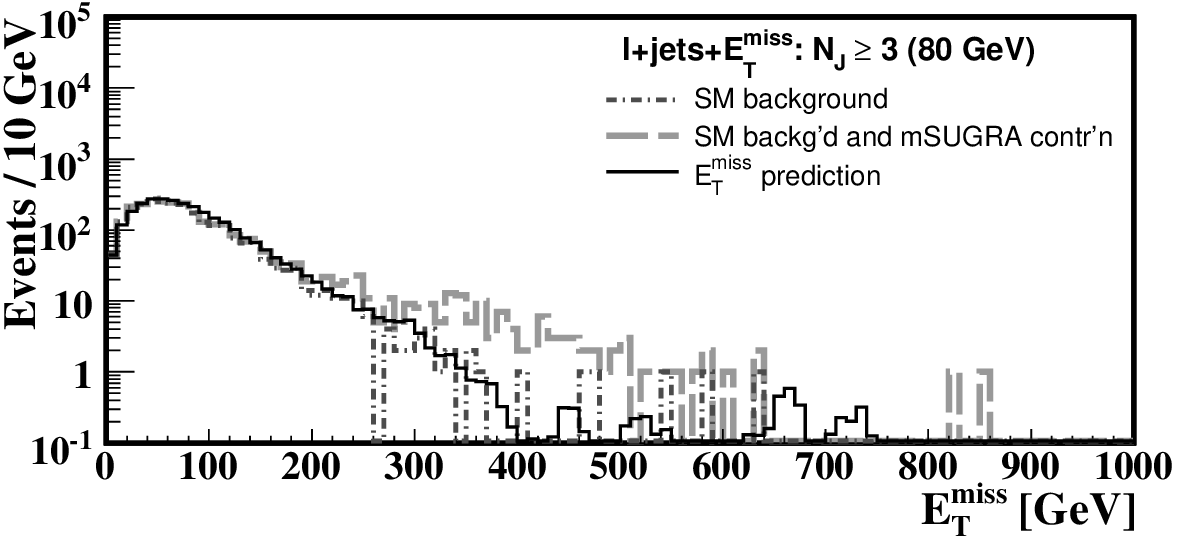,width=3.5in}
\epsfig{file=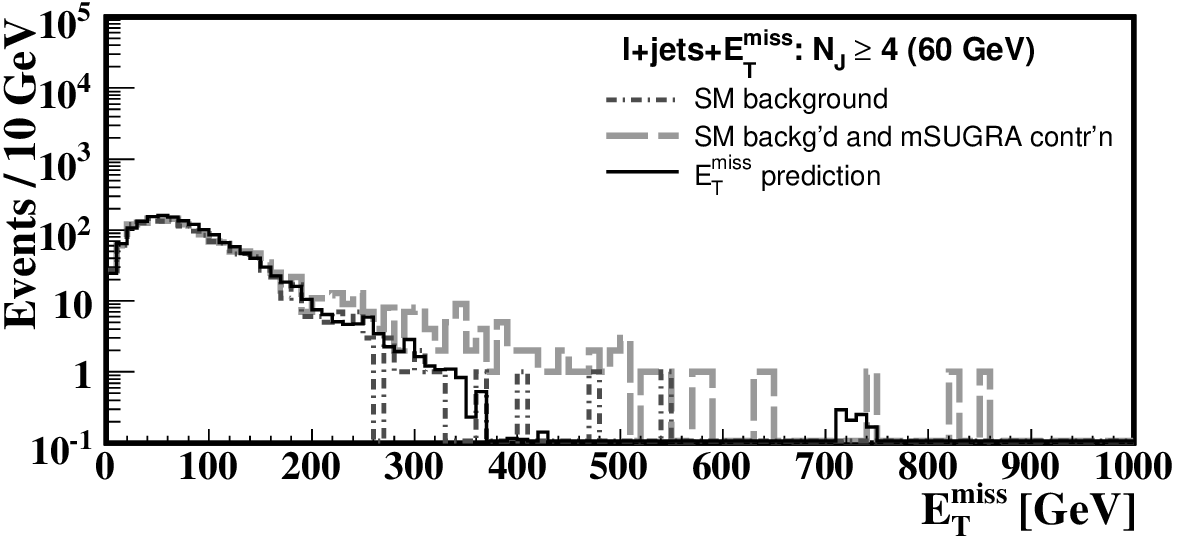,width=3.5in}
\end{center}
\end{minipage}
\end{center}
\caption{ Observed~(dashed) and predicted~(solid) SM \zjets~(top) and \ljetsmet~(bottom) 
          for $N_J \ge 3$~(left) and $N_J \ge 4$~(right) with new physics contributions 
          from mSUGRA benchmarks~\cite{{cms-msugra}}. 
	  The dot-dashed lines highlight the SM contributions. 
          The plots are made for the high \pt~ thresholds for $N_J$ and 
          correspond to 200~pb$^{-1}$ at $\sqrt{s} = 14$~TeV.
	}
\label{figure-13} 
\end{figure*}

The algorithm's performance with a new physics contribution 
is illustrated in Figure~\ref{figure-12} in the  \zjets~ and  
\ljetsmet~ channels in events with $N_J = 3$~(left) and~$\ge 4$~(right) 
for the 50~GeV threshold.  Figure~\ref{figure-13} shows the 
corresponding distributions in events with $N_J \ge 3$~(left) 
and~$\ge 4$~(right) for the high \pt~ jet thresholds.
The integrated luminosity of the mock data samples in these 
Figures is $200~{\rm pb}^{-1}$ for $\sqrt{s} = 14$~TeV.
New physics contributions in the Figures are similar to 
those from mSUGRA benchmarks LM4 and LM1~\cite{cms-msugra} 
for \zjets~ and \ljetsmet, respectively. 
The plots show SM backgrounds with new physics 
contributions~(dashed) and their \met~ predictions~(solid) 
from QCD templates. The dot-dashed lines represent 
SM backgrounds only to ease comparisons.

New physics events tend to have large $J_T$ and \met.
It is seen that the addition of a signal contribution 
with large $J_T$ does not bias the prediction 
significantly at high \met. An excess of signal events above the 
background prediction stands out clearly in both channels. 
Since in \ljetsmet~ the neutrino spectrum is modeled 
based on the charged lepton spectrum in each $N_J$ bin, 
the method works best in this signature when the charged 
lepton spectrum in new physics events is soft compared 
to the \pt~ spectrum produced by new weakly interacting 
particles~\cite{spectra-consistency}.

\section{Conclusion}

\label{summary}

I have presented a new method to predict SM backgrounds at high \met~ 
and a large number of jets, $N_J$,
within a context of a search for new phenomena in final states consistent
with SM \vjets: \zjets, \gammajets~ and \wjets. 
The artificial \met~ in each \vjets~ event is modeled 
in-situ using multi-jet QCD events with a configuration of jets similar 
to that in the \vjets~ event. The genuine \met~ contribution from 
neutrinos in the \ljetsmet~ channel, dominated by \wjets~ and \ttbarjets, 
is modeled based on the charged lepton \pt~ spectrum.

The method performs reasonably well in robustness tests. 
I have identified mechanisms by which it may become biased, 
discussed systematic uncertainties in its background predictions 
and procedures to estimate them. 
A new physics contamination of the QCD sample does not lead to 
a significant bias. The method has discriminating power to reveal 
a new physics contribution at high \met~ and a large number of jets. 
It can be applied to data with minimal recourse to MC simulation 
in early LHC running when robust data-driven SM background 
predictions play a key role in searches for new phenomena.


\begin{thebibliography}{99}
\bibitem{ttbar}

The \wjets~ and \ttbarjets~ processes are studied in the \ljetsmet~ 
signature and treated in the same manner in the paper. 
In a few places, where differences between the two processes are 
important, they are discussed separately.
\bibitem{eta}

Pseudo-rapidity is $\eta = -\ln[ \tan(\theta/2)]$,
where $\theta$ is the particle's polar angle to the beam line.
\bibitem{elec-jet-ambiguity} 

Energy depositions in the calorimeters from electrons~[$Z(ee)$+jets~ 
and $e$+jets+\met] and photons~(\gammajets) are reconstructed as jets.  
Such jets are well-measured since most of their energy is collected by 
the electromagnetic calorimeter.
In order to avoid double counting the energy depositions from electrons 
and photons in \met, only jets  are used in \met~ for final states 
with electrons or photons. 
In $J_T$ and $N_J$ measurements, jets from electrons and photons
are not used.
\bibitem{jet-loss} This is a simple model of jet losses that allows to 
study how the method performs if there are other sources of  jet inefficiency,
for example, more uniform in $\eta$ and $\phi$.
\bibitem{b-jets} 

The contribution from $b$-jets to the bias in artificial \met~
prediction is smaller than that due to differences in $N_J$ and $J_T$ spectra
between \ttbarjets~ and QCD. 
Resolution effects specific to higher track multiplicity $b$-jets are not 
simulated in this study.
\bibitem{w-polarn} The effect of $W$-polarization in \wjets~ on the charged 
lepton and neutrino \pt~ spectra are more pronounced in the forward regions 
at high $|\eta_{\rm charged\; lepton}|$, especially outside the fiducial 
$\eta$ coverage for charged leptons. Therefore, they can be enhanced 
or reduced by varying the $\eta_{\rm charged\; lepton}$ selection.
\bibitem{spectra-consistency}

More generally, in \ljetsmet, one could develop a search 
for new physics by studying the consistency of 
the charged lepton \pt~ spectrum with that of the neutrino.
\end{thebibliography}
\end{document}